\documentclass[onecolumn]{article}
\usepackage{graphicx}
\usepackage{authblk}
\usepackage[left=2cm,right=2cm]{geometry}
\usepackage{amsmath, amscd, amssymb, amsthm}
\usepackage[final]{microtype}

\usepackage{url}
\usepackage{caption}
\usepackage{subcaption}
\usepackage[title]{appendix}
\usepackage{multirow}

\usepackage{titlesec}
\titleformat{\section}[hang]{\bfseries\large}{\thesection}{1em}{\raggedright}

\title{Spacecraft heat shield study in the DIII-D tokamak}

\author[1]{Dmitri M. Orlov}
\author[2]{Evdokiya G. Kostadinova}
\author[3]{Igor Bykov}
\author[1]{Dmitri L. Rudakov}
\author[1]{Roman Smirnov}
\author[3]{Jayson Barr}
\author[2]{Gabrielle Bladon}
\author[8]{Alessandro Bortolon}
\author[1]{Justin Burzachiello}
\author[3]{Lane Carlsson}
\author[3]{Colin Chrystal}
\author[1]{Jason Escalera}
\author[2]{Jessica Eskew}
\author[9]{Graeson Griffin}
\author[1]{Michael O. Hanson}
\author[10]{Georg Herdrich}
\author[6]{Jeffrey Herfindal}
\author[3]{Al Hyatt}
\author[9]{Truell Hyde}
\author[5]{Charles Lasnier}
\author[1]{Claudio Marini}
\author[9]{Lorin Matthews}
\author[5]{Adam McLean}
\author[2]{Christopher A. Mehta}
\author[1]{Renato Perillo}
\author[4]{Jens Schmidt}
\author[5]{Filipo Scotti}
\author[2]{Zola Spence}
\author[1]{Hadith Taheri}
\author[3]{Michael van~Zeeland}
\author[1]{Caitlyn Villareal}
\author[3]{Huiqian Wang}
\author[6]{Robert Wilcox}
\author[7]{Theresa Wilks}
\author[11]{Nandini Yadav}
\author[1]{Daniel Zubovic}

\affil[1]{University of California San Diego, La Jolla, CA 92093-0417, USA}
\affil[2]{Auburn University, Auburn, AL 36849, USA}
\affil[3]{General Atomics, San Diego, CA 92186, USA}
\affil[4]{DLR, Germany}
\affil[5]{Lawrence Livermore National Laboratory, Livermore, CA 94550, USA}
\affil[6]{Oak Ridge National Laboratory, Oak Ridge, TN 37831, USA}
\affil[7]{Massachusetts Institute of Technology, Cambridge, MA 02139, USA}
\affil[8]{Princeton Plasma Physics Laboratory, Princeton, NJ 08543, USA}
\affil[9]{Baylor University, Waco, TX 76798, USA}
\affil[10]{University of Stuttgart, 70569 Stuttgart, Germany}
\affil[11]{Oak Ridge Associate Universities, Oak Ridge, TN 37831, USA}

\date{March 2026}

\begin{document}


\twocolumn[
  \begin{@twocolumnfalse}
    \maketitle
    \begin{abstract}
  We report on a new experimental platform developed at the DIII-D National Fusion Facility to investigate carbon ablation and spallation under extreme heat flux conditions relevant to both plasma-facing components in fusion facilities and high-enthalpy atmospheric entries. Carbon samples were exposed to parallel heat fluxes of $30\text{--}40~\mathrm{MW\,m^{-2}}$ in the scrape-off layer plasma using two complementary approaches: (1) stationary carbon rods of varied geometric profiles inserted near the divertor strike point, and (2) slow-launch carbon pellets injected vertically into the edge and core plasma from the DiMES port. For pellets penetrating the core plasma, the heat flux was an order of magnitude higher. These conditions were selected to reproduce key aspects of the shock-layer heating environment experienced by the Galileo probe during its entry into Jupiter’s atmosphere, particularly the combination of high parallel heat flux and strong plasma flows. Visible fast-frame imaging, divertor spectroscopy, infrared thermography, CO$_2$ interferometry interferometry, and post-exposure surface profilometry enable direct measurement of ablation rates, surface recession patterns, and surface temperature evolution of the exposed samples. The measured mass loss rates of $(1\text{--}3)\times10^{-2}~\mathrm{g\,cm^{-2}\,s^{-1}}$ are consistent with semi-empirical ablation models developed in the aerospace community, with wedge rod geometries exhibiting enhanced ablation relative to cylindrical and concave geometries. UEDGE-DUSTT simulations incorporating strong parallel flows, $\mathbf{j}\times\mathbf{B}$ forces, and ablation-cloud shielding reproduce the measured pellet trajectories and ablation timescales. These results establish tokamak plasma as a unique high-heat-flux test environment capable of validating ablation models for carbon-based protective materials while simultaneously advancing understanding of material response and impurity source dynamics in reactor-relevant divertor plasmas.    
    \end{abstract}
  \end{@twocolumnfalse}
]

\section{Introduction}

Atmospheric entries of satellites and probes present extreme heating environments, especially entries into planets with thick atmospheres. The mitigation of peak heating through the use of blunt-body geometries and detached shock layers is a foundational result of hypersonic entry theory \cite{AllenEggers1958}. NASA entry probes have successfully navigated conditions ranging from mild to severe. For example, during the Mars Viking atmospheric entry, peak heat fluxes reached around 0.25~MW\,m$^{-2}$, with a stagnation pressure of approximately 0.05~atm \cite{Cutts2005EntryProbes}. The Apollo and \textit{Stardust} missions encountered peak heat fluxes of 5--10~MW\,m$^{-2}$ and stagnation pressures between 0.3 and 1~atm during reentry into Earth's atmosphere \cite{Tauber1991Apollo,AllenEggers1958,Liu2010StardustRadiation}. The Pioneer Venus probe experienced even higher peak heat fluxes of 100~MW\,m$^{-2}$ and stagnation pressures of 6--7~atm \cite{Moss1982,CabreraWest2025PioneerVenus}. For over 60~years, ablative materials have effectively protected valuable research equipment on descent modules and probes in such extreme conditions \cite{LaubVenkatapathy2003}. 

The selection of carbonaceous thermal protection materials depends heavily on the anticipated entry environment. Low-density architectures (such as porous carbon matrices) serve as effective thermal insulators in moderate radiative environments, whereas high-density carbon composites and graphites are required as effective ablators to sustain extreme heat fluxes and mechanical shear without undergoing premature spallation \cite{LaubVenkatapathy2003}. In the context of spacecraft entry, controlled ablation is a process in which the heat shield material dissociates while absorbing energy, thus preventing the remainder of the spacecraft from overheating. However, beyond a threshold heat flux incident on the material, excessive mass loss can occur due to spallation - a process in which whole fragments can be ejected from the material surface. Spallation is dangerous for the spacecraft as it can remove the heat shield material too fast and alter the probe's shape, affecting its aerodynamic stability. To account for such unwanted processes, it is common to allow for large safety margins when building the heat shield, which substantially increases its mass. Minimizing the weight of the thermal protection system (TPS) by constraining safety margin requirements ensures more space for the scientific payload on these missions. Improved predictions of mass loss rates and understanding the failure modes of TPS materials, such as the onset of spallation, are crucial for the improvement of heat shields needed for future missions, especially ones that include probe entries into the gas giants.

One of the most extreme atmospheric entries occurred on 7 December 1995, when the Galileo probe entered Jupiter's atmosphere. Descending at a relative velocity of 47.4~km\,s$^{-1}$ (or 29.5 ~mi/s), the probe experienced a peak heating rate of 300~MW\,m$^{-2}$, with heat loads of 3~GJ\,m$^{-2}$, while shock layer temperatures peaked at around 16{,}000~K \cite{Moss1982, matsuyama2005numerical}. Approximately 26.5\% of the probe's initial mass of 335~kg was lost during the 70~s heat pulse \cite{Milos1999}. While thermochemical ablation of the forebody heat shield accounted for a loss of 79~kg (approximately 23.6\% of the total vehicle mass), the remaining 2.9\% of the initial mass is attributed to other high-stress mass-loss mechanisms, such as mechanical spallation.

The Galileo probe's heat shield was an axisymmetric sphere-cone design with a 22.2~cm nose radius and a $44.86^\circ$ half-angle frustum. Its composition included chopped-molded carbon phenolic in the nose cap and tape-wrapped carbon phenolic along the frustum, a configuration initially developed for the Pioneer Venus probe. The material had a virgin density of $\rho_\mathrm{v} = 1448~\mathrm{kg\,m^{-3}}$ and a char density of $\rho_\mathrm{c} = 1185~\mathrm{kg\,m^{-3}}$\cite{Milos1999}. Extensive pre-flight experimental studies, including laser irradiation, ballistic range, and plasma arc-jet tests \cite{Moss1982}, informed the design of the heat shield. During entry, 10 analog resistance ablation detector (ARAD) sensors embedded in the forebody provided real-time data on the recession of the heat shield, which was transmitted to the orbiter \cite{Milos1999}. This data allowed scientists to compare actual recession rates to predicted values at various locations of the heat shield, including the stagnation point and downstream regions.

Post-flight analysis revealed substantial discrepancies between ARAD measurements and the predictive models used in the design. The models overestimated ablation at the stagnation point while underestimating it in downstream regions. The results from Galileo's atmospheric entry highlighted both challenges, as only 1~cm of the initial 5.4~cm thickness remained in the downstream region by the end of entry, while the stagnation point retained 10~cm of the original 14.6~cm thickness \cite{Milos1999, matsuyama2005numerical}. The Galileo probe survived its mission due to a very high engineering safety margin. Designing improved heat shields requires balancing the shield's mass to reduce launch costs and ensuring adequate protection for onboard equipment.

The discrepancies between predicted and measured mass loss rates for the Galileo probe inspired the development of improved ablation models for high heat flux environments, notably by researchers such as Park and Matsuyama \cite{matsuyama2005numerical, park2009stagnation}. In this work, we aim to validate the predictions of these models for a range of experimental heating conditions achieved in the DIII-D tokamak. 

Another key conclusion from the Galileo mission was the limitation of laboratory tests in fully replicating the conditions of high-speed atmospheric entry, including high heat flux, boundary layer dynamics, heating history, and spallation effects. In this work, we present for the first time how conditions in the edge of a tokamak plasma can closely replicate those of high-speed atmospheric entry. We also demonstrate how tokamak plasmas can be utilized to test heat shield materials and validate models that can guide the design of next-generation spacecraft thermal protection systems. It should be noted that, while the tokamak environment closely replicates the plasma temperature, density, and heat flux of a gas giant entry shock layer, it does not simulate the high stagnation pressures (up to 7 atm) encountered during flight. Consequently, the focus of this platform is the validation of thermochemical ablation, ionization effects, and plasma-surface interactions rather than mechanical pressure loading.

The remainder of this paper is organized as follows. Section~\ref{sec:tokamak} describes the DIII-D experimental environment, its similarity to atmospheric-entry conditions, and the diagnostics that enable quantitative measurements. Section~\ref{sec:ablation_models} summarizes the semi-empirical models by Park and Matsuyama that form the basis for comparison between atmospheric entry predictions and the DIII-D experiment. Section~\ref{sec:rods} presents the results from experiments where stationary rods were exposed to edge plasma. This section discusses heat-flux reconstruction, spectroscopy, camera imaging, and post-mortem profilometry used to infer mass-loss rates for the rod ablation experiments. Section~\ref{sec:pellets} presents results where pellets were injected deeper in the tokamak plasma. This section discusses reconstruction of pellet trajectories, ablation dynamics, interferometry-based mass estimates, and UEDGE--DUSTT modeling, and reports observations of spallation. Finally, Section~\ref{sec:conclusions} summarizes the principal results and discusses implications for validation of ablation physics in both fusion and atmospheric-entry environments.
\section{Use of tokamaks to study carbon ablation in high heat flux plasma conditions}
\label{sec:tokamak}

The experiments presented in this study were conducted at the DIII-D National Fusion Facility \cite{Luxon2002} in San Diego, CA, as part of the Frontier Plasma Science campaign. DIII-D is a medium-sized tokamak with a major radius of 1.67~m and a minor radius of 0.67~m, capable of sustaining plasma discharges for 5--7~s. With an all-carbon wall design, DIII-D is particularly well-suited for investigating carbon ablation, a critical process in understanding heat shield behavior in high-heat-flux environments.

\begin{figure}[htbp]
    \centering

    \begin{subfigure}[b]{0.3\textwidth}
        \centering
        \includegraphics[width=\textwidth]{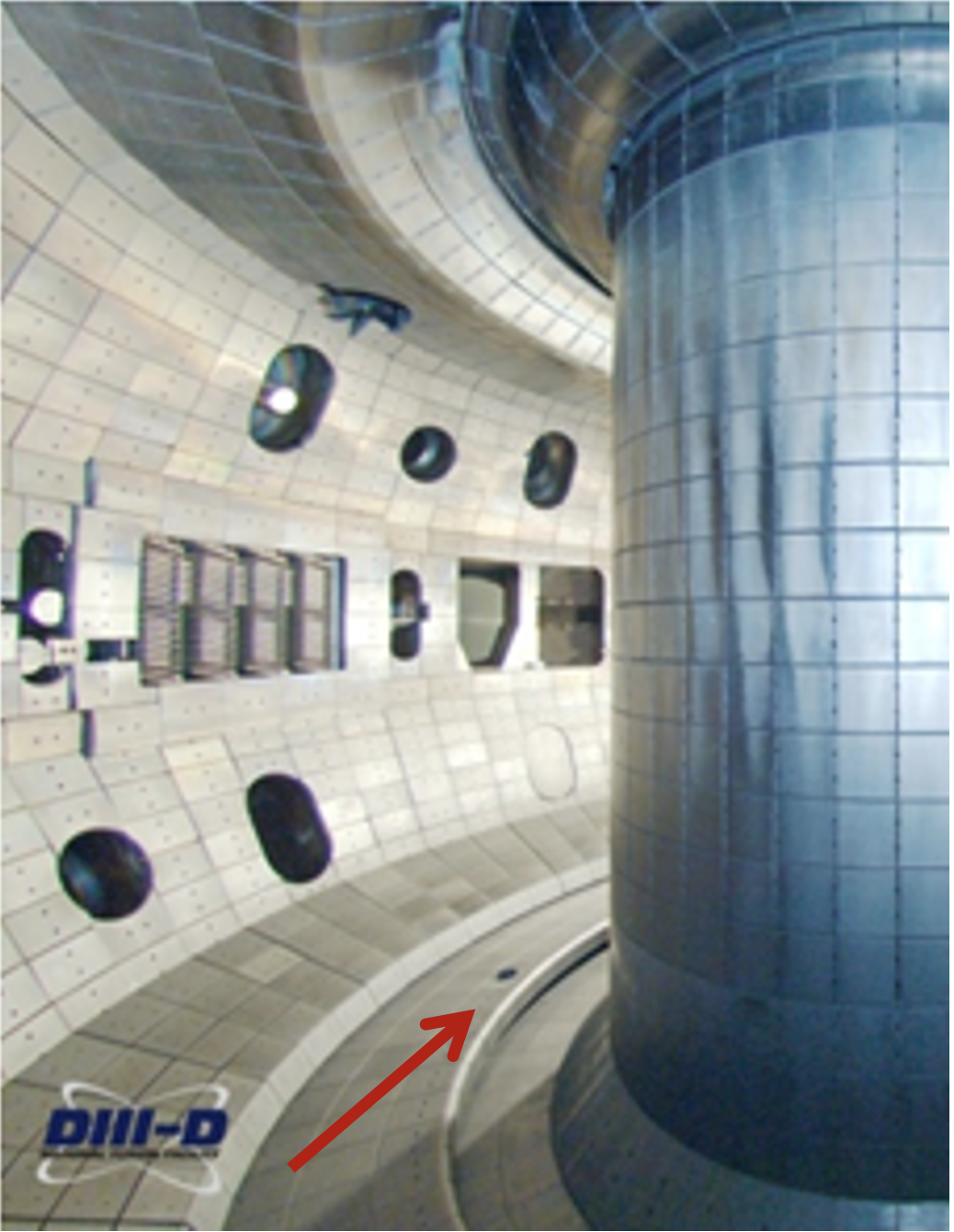}
        \caption{Lower divertor region of DIII-D showing the DiMES port (red arrow).}
        \label{fig:f1a}
    \end{subfigure}
    \hfill
    \begin{subfigure}[b]{0.3\textwidth}
        \centering
        \includegraphics[width=\textwidth]{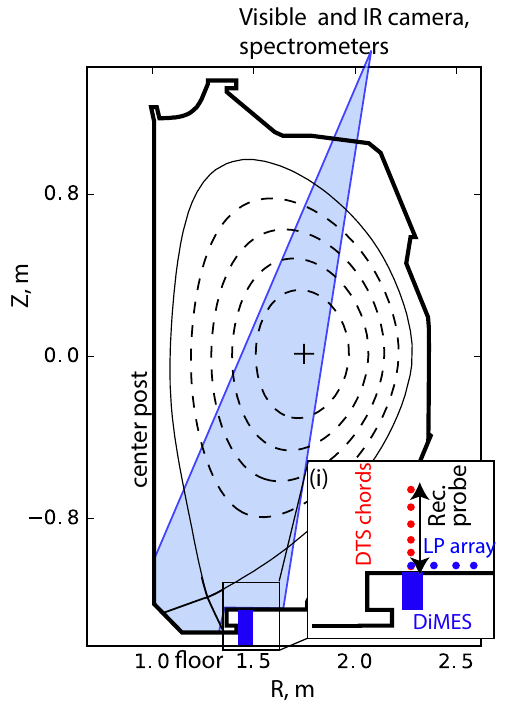}
        \caption{Poloidal cross-section showing DiMES location and several diagnostic sight lines: Divertor Thomson Scattering (DTS) chords (red dots) and Langmuir probe (LP) array (blue dots). Blue shaded region shows the sight line for visible and infrared cameras.}
        \label{fig:f1b}
    \end{subfigure}

    \caption{DiMES port in the lower divertor of DIII-D and sight lines of several key diagnostics.}
    \label{fig:fig1}
\end{figure}

The Divertor Material Evaluation System (DiMES)  is designed for exposing material samples to divertor plasma under well-characterized heat flux conditions \cite{Wong1998DiMES}. In standard operation, samples are mounted level with the divertor graphite tile surface, enabling controlled exposure with fixed strike point location, as well as the outer strike point sweep across the DiMES location \cite{Rudakov2017} for discharge conditions characterization. For the present work, the DiMES hardware was modified to position ablating carbon targets directly in the power exhaust region of the scrape-off layer, as indicated in Fig.~\ref{fig:fig1} (a, b). Two complementary ablation techniques are utilized: (1) partial ablation of carbon rods extending into the scrape-off layer (SOL) from the DiMES port and (2) full ablation of spherical carbon pellets launched vertically into the plasma from the same location. The rods were made of ATJ graphite, while the pellets were either made of porous or glassy carbon, thus allowing the investigation of various carbonaceous structures. These two approaches, described in Secs.~IV and~V, enable direct measurement of ablation rates and plasma-driven shape evolution under entry-relevant parallel heat fluxes.

\begin{figure}[htbp]
    \centering
    \includegraphics[width=0.47\textwidth]{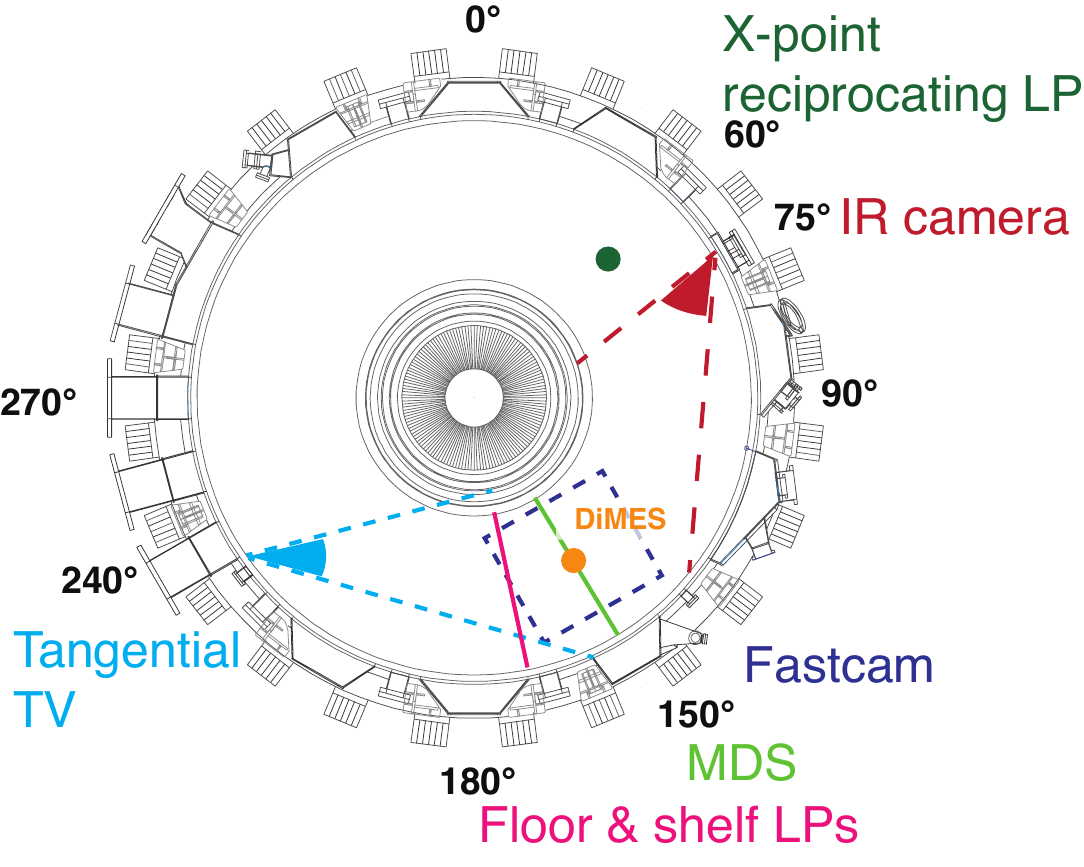}
    \caption{Schematic toroidal cross-section showing the locations of diagnostics near the DiMES port in DIII-D; the drawing is not to scale. Visible-imaging systems, including the Fastcam and tangential TV cameras, and IR cameras provide information on the sample location, temperature, and carbon emission. The Multichordal Divertor Spectrometer (MDS) viewing chords provide additional measurements of carbon emission, while Langmuir probes and Thomson scattering viewing chords (not shown) provide information on the plasma temperature and density in the region surrounding the ablating samples.}
    \label{fig:dimes}
\end{figure}

Figure~\ref{fig:dimes} shows a toroidal cross-section of the tokamak and the DiMES port location in the lower divertor along with the field of view of several key diagnostics. A visible imaging fast-framing camera views the DiMES location from above \cite{Moyer2018RSI}, and an infrared (IR) periscope \cite{Lasnier2014RSI} provides a tangential view from the outboard midplane side. The Multichordal Divertor Spectrometer (MDS) \cite{brooks_howald_klepper_west_1992} has a dedicated chord viewing the DiMES location directly, enabling spectroscopic measurements of carbon line and molecular band emission. Additional diagnostics include the divertor Thomson scattering (DTS) system \cite{Carlstrom1995DTS}, a reciprocating Langmuir probe \cite{Watkins1992RecipProbe}, and fixed floor Langmuir probes \cite{Buchenauer1990FixedProbes}, which provide information on electron temperature, density, and heat flux near the DiMES region.

The plasma discharge parameters in DIII-D’s scrape-off layer (SOL) and divertor region near the DiMES port enable the simulation of high heat flux environments comparable to those experienced by the Galileo probe during its entry into Jupiter’s atmosphere. A comparison of key parameters is provided in Table~\ref{tab:comparison}. Jupiter’s atmosphere is primarily composed of hydrogen (86.4\%) and helium (13.6\%), while DIII-D typically operates with deuterium as the main gas, sometimes with hydrogen or helium seeding. Although hydrogen and deuterium differ in ion mass, we expect that this difference has only a minor influence on heat flux characteristics.



Galileo entry conditions featured peak plasma temperatures of approximately 1.4~eV ($\approx 16{,}000$~K) \cite{Moss1982, matsuyama2005numerical}, which approach the lower limits of the 10--100~eV range characteristic of the DIII-D edge plasma. While the upstream Jovian atmosphere is predominantly neutral and several orders of magnitude lower in ionized particle density than a tokamak plasma, the hypersonic probe velocity ($\sim 50$~km\,s$^{-1}$) generates a strong detached shock ahead of the heat shield. Rapid dissociation, ionization, and compression across this shock produce a dense post-shock plasma with particle densities in the $10^{18}$--$10^{20}$~m$^{-3}$ range, similar to those encountered in the DIII-D scrape-off layer \cite{Tauber1999NASA, Erb2020GalileoHeating}.

In addition, the incident heat fluxes of $100$--$500$~MW\, m$^{-2}$ experienced by the Galileo probe overlap with parallel heat fluxes, $q_{\parallel}$, $50$--$200$~MW\,m$^{-2}$ heat fluxes achievable at the DiMES location. Ion flow velocities parallel to the magnetic field in DIII-D ($10$--$80\text{~km~s}^{-1}$) are also comparable to the relative entry velocity of the Galileo probe.
Finally, the duration of a typical DIII-D plasma discharge is $5-6~s$, which reproduces more closely the expected heating history when compared to other laboratory tests, such as material ablation with nanosecond lasers. These similarities establish tokamak experiments as a unique high-enthalpy testbed for evaluating ablation physics relevant to atmospheric entries.

\begin{table}[t]
\centering
\footnotesize
\setlength{\tabcolsep}{3pt}
\caption{Comparison between Galileo probe entry conditions in the shock layer plasma and plasma conditions accessible in the DIII-D scrape-off layer (SOL).}
\label{tab:comparison}
\begin{tabular}{lcc}
\hline\hline
\textbf{Parameter} & \textbf{Galileo--Jupiter} & \textbf{DIII-D tokamak} \\
\hline
Gas composition &
H$_2$ (86.4\%) &
D \\
& He (13.6\%) & H, He \\[2pt]

Temperature &
$\lesssim 100$~eV &
12--100~eV \\[2pt]

Density\footnotemark[1] &
$\lesssim 10^9$~m$^{-3}$ &
$10^{18}$--$10^{19}$~m$^{-3}$ \\[2pt]

Velocity &
$\approx 50$~km\,s$^{-1}$ &
10--80~km\,s$^{-1}$ \\
& (entry) & (ion flow) \\[2pt]

Heat flux &
100--500~MW\,m$^{-2}$ &
50--200~MW\,m$^{-2}$ \\
\hline\hline
\end{tabular}
\end{table}

\footnotetext[1]{Behind the hypersonic shock formed ahead of the probe, the post-shock plasma density becomes comparable to DIII-D edge plasma densities (see text) \cite{Tauber1999NASA, Erb2020GalileoHeating}.}

This unique combination of high heat flux, strong parallel plasma flows, advanced divertor diagnostics, and controlled exposure geometry makes tokamak plasmas a valuable platform for the study of ablation physics relevant to both fusion reactors and high-enthalpy aerospace applications.

Ground-based arc-jet testing and post-flight analysis of the Galileo probe TPS data have provided essential insights into carbon ablation under high-enthalpy atmospheric entry conditions. However, the interpretation of the in-flight ablation data remains complicated by uncertainties in the radiation environment, ablation-cloud shielding, and strong ionization effects in the Jovian entry plasma. These uncertainties lead to significant model-to-model variation in predicted ablation rates and TPS mass margins for current and future planetary entry missions.

To improve the predictive capability of carbon ablation models, laboratory experimental data are needed in environments where heat flux, ion flow, and plasma radiation can be controlled and diagnosed with high precision. The carbon ablation studies presented in this work provide such a platform by leveraging the DIII-D tokamak edge plasma as a unique high-heat-flux test environment. By combining direct measurement of carbon mass loss rates, stereoscopic detection of ablation products, multi-instrument measurements of the plasma conditions, and supporting numerical modeling, these experiments supply the physics constraints required for validating and refining carbon ablation models developed using both arc-jet testing and Galileo mission data.

The DIII-D results reported here form a critical connection between traditional TPS development pathways and new plasma-based ablation validation capabilities, enabling more accurate modeling of carbon-based heat-shield performance in extreme planetary-entry environments.
\section{Semi-empirical models for carbon ablation}
\label{sec:ablation_models}

In this section, we discuss two semi-empirical models of carbon mass loss rates proposed in the aerospace community. We use these models to predict mass loss rates for the stationary rod experiments in DIII-D, for which steady-state plasma conditions can be assumed. We then compare the predictions against the measured mass loss rates from these experiments and discuss validity of each model. For the pellet experiments, steady-state plasma conditions cannot be assumed as the pellets move from the SOL to the core plasma where they fully ablate. Instead, the results from these experiments are compared against simulations using the UEDGE-DUSTT code, as discussed in Sec.~\ref{subsec:data_workflow}.

Here we assume that the ablation is quasi-steady-state for either an atmospheric entry or under high heat flux testing in a tokamak or other laboratory facility. In this case, the surface mass loss rate $\dot{m}$ (in kg\,m$^{-2}$\,s$^{-1}$) can be determined from the energy balance at the wall using the following expression  \cite{matsuyama2005numerical}
\begin{equation}
    \dot{m} = \frac{-(q_{rw} + q_{cw})}{\Delta H_a}.
    \label{eq:massloss_general}
\end{equation}
Here, $q_{rw}$ (MW\,m$^{-2}$) and $q_{cw}$ (MW\,m$^{-2}$) are the radiative and convective heat transfer rates to the wall, respectively, and $\Delta H_a$ (MJ\,kg$^{-1}$) is the heat of ablation, or effective energy needed to remove a unit mass of material from the surface. In atmospheric entry conditions, the convective heat flux arises due to the direct contact of the heat shield and the hot compressed gas, while radiative heat flux results from electromagnetic radiation due to atomic processes (ionization, excitation, etc.) in the shock layer. In DIII-D, a common measure of heat flux is the "parallel heat flux" $q_{\parallel}$, or the heat flux delivered to the surface by the plasma particles streaming along the magnetic field lines. Since this mechanism for delivery of heat to the surface is based on contact between the plasma particles and the target surface, it is reminiscent to convective heat flux with the important distinction that for $q_{\parallel}$, the energy is preferentially moved along the direction of the magnetic field. Some radiative heat flux due to atomic processes in the DIII-D SOL plasma is expected, but a much smaller effect than the parallel heat flux. 

Here we calculate $q_{\parallel}$ from measured plasma conditions according to \cite{Watkins2001JNM}
\begin{equation}
    q_{\parallel}=\gamma_ek_BT_en_ec_s,
\end{equation}
where $k_B$ is the Boltzmann constant, $T_e$ and $n_e$ are the electron temperature and density, respectively, $c_s=(2k_BT_e/m_i)^{1/2}$ is the plasma sound speed (assuming equal electron and ion temperatures, $m_i$ is the ion mass), and $\gamma_e$ is the sheath power transmission factor (SPTF), which is empirically obtained for DIII-D divertor plasma conditions. The SPTF factor is known to vary in the range $\gamma_e\approx7-9.5$ (see Fig 6 from \cite{donovan2013experimental}). In this work, we assume a value of $\gamma_e=7$. 

In this paper, we calculate mass loss rates using a value for $q_{\parallel}$ estimated from experimentally measured electron densities and temperatures from Thomson scattering and Langmuir probes. The heat of ablation $\Delta H_a$ is calculated using two different semi-empirical models developed for carbon ablation during the Galileo entry into Jupiter. The first model by Matsuyama~\cite{matsuyama2005numerical} proposes that the heat of ablation in units of $MJ~kg^{-1}$ for carbon-based ablators could be related to the surface (or wall) pressure $p_w$ as
\begin{equation}
    \Delta H_a = 28 - 1.375\,log~p_w + 27.2\,(log~p_w)^2,
    \label{eq:matsuyama}
\end{equation}
where $p_w$ is the surface pressure in atmospheres and can be related to the vapor pressure over the ablating surface. Here, we take $p_w$ to be equal to the saturated vapor pressure $p_v$. For graphite, a semi-empirical relation for the dependence of $p_v$ on surface (or wall) temperature $T_w$ is given by \cite{Smirnov2007}

\begin{equation}
    p_v=10^{14.8-40181/T_w}.
    \label{eq:vap_pressure}
\end{equation}

Equation~\eqref{eq:vap_pressure} yields pressure in units of $Pa$ and should be multiplied by a factor of $9.86923\times10^{-6}$ to obtain units of $atm$ before plugging in Eq.~\eqref{eq:matsuyama}. 

The second model examined here is by Park~\cite{park2009stagnation}. The model relates the heat of ablation to the surface (or wall) temperature $T_w$ according to
\begin{equation}
    \Delta H_a = 23307 - 0.825\,(T_w - 4000),
    \label{eq:park}
\end{equation}
where $T_w$ is expressed in $K$ and $\Delta H_a$ is in J\,g$^{-1}$. This expression is expected to be valid for temperature range $(3700-4100)~K$. 

To provide estimates of the ablation rates predicted from each model, we assume that the wall temperature $T_{w}$ is equal to the sublimation temperature $T_{s}$ of the material, and that the sublimation temperature assumed for the ATJ graphite used in our rod experiments is $3800~K$. Considering the measured incident parallel heat flux was in the range $q_{\parallel} = 30$–$40$~MW\,m$^{-2}$ (see Fig. 4), the Matsuyama model, Eq.~\eqref{eq:matsuyama}, predicts mass loss rates on the order of $(0.66$–$0.87)$~kg\,m$^{-2}$\,s$^{-1}$, while the Park model, Eq.~\eqref{eq:park}, yields mass loss rates of $(1.28$–$1.70)$~kg\,m$^{-2}$\,s$^{-1}$. Both estimates predict measurable mass loss rates for $cm$-sized rod samples exposed to the DIII-D SOL plasma for $\approx5-6~s$ per shot. However, the mass loss rates predicted by the Park model are roughly a factor of $2$ larger than those predicted with the Matsuyama model. Such differences should be within the precision of the mass loss measurements achievable in the DIII-D experiments.
\section{Ablation of stationary carbon rods}
\label{sec:rods}



\begin{figure}[htbp]
    \centering
    \begin{subfigure}[b]{0.47\textwidth}
        \centering
        \includegraphics[width=\textwidth]{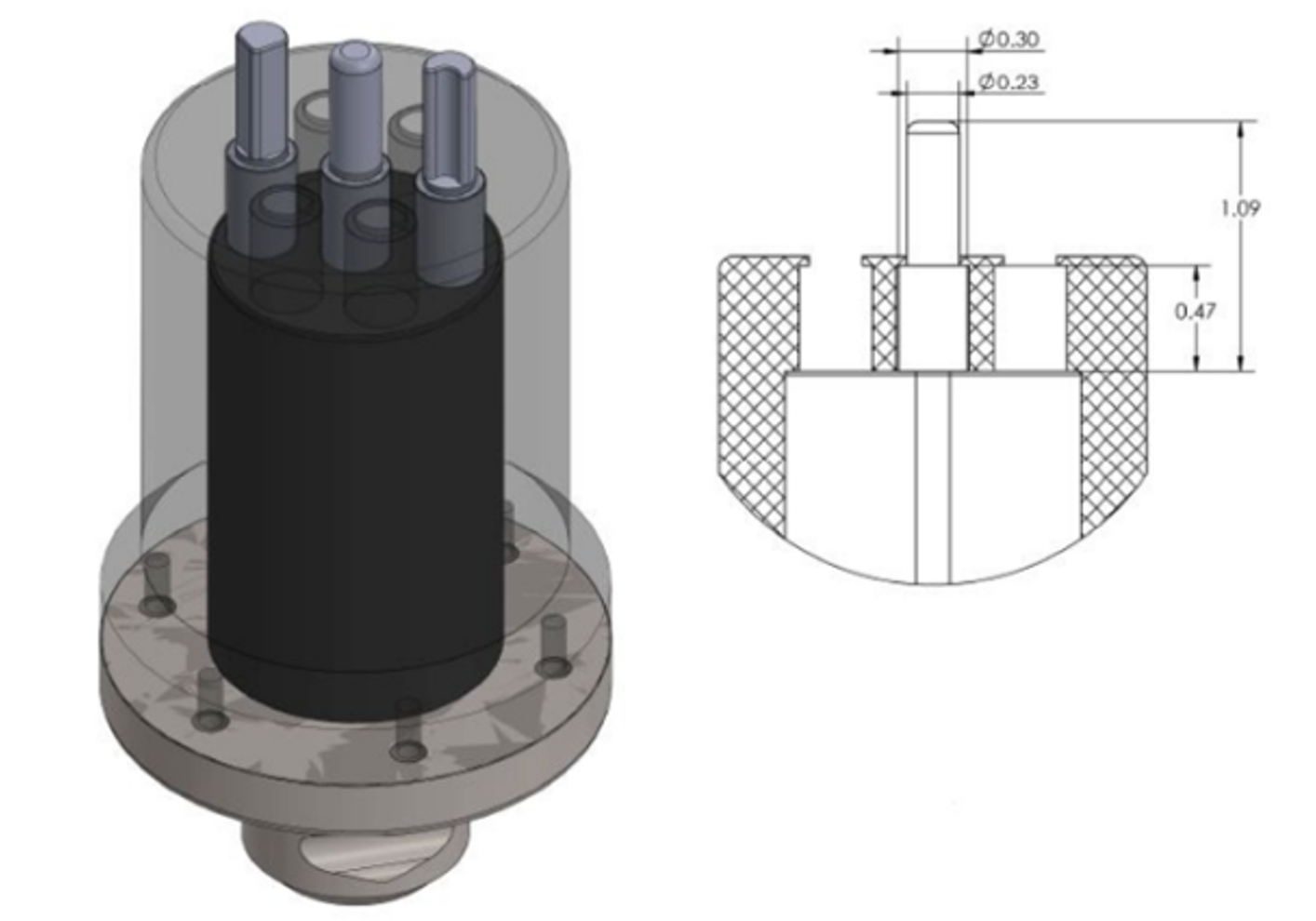}
        \caption{Example configuration of the modified DiMES head with three carbon ablation rods. The specific combination of rod shapes shown here is illustrative; different combinations of cylindrical, wedge, and concave rods were used during the experiments.}
    \end{subfigure}
    \hfill
    \begin{subfigure}[b]{0.47\textwidth}
        \centering
        \includegraphics[width=\textwidth]{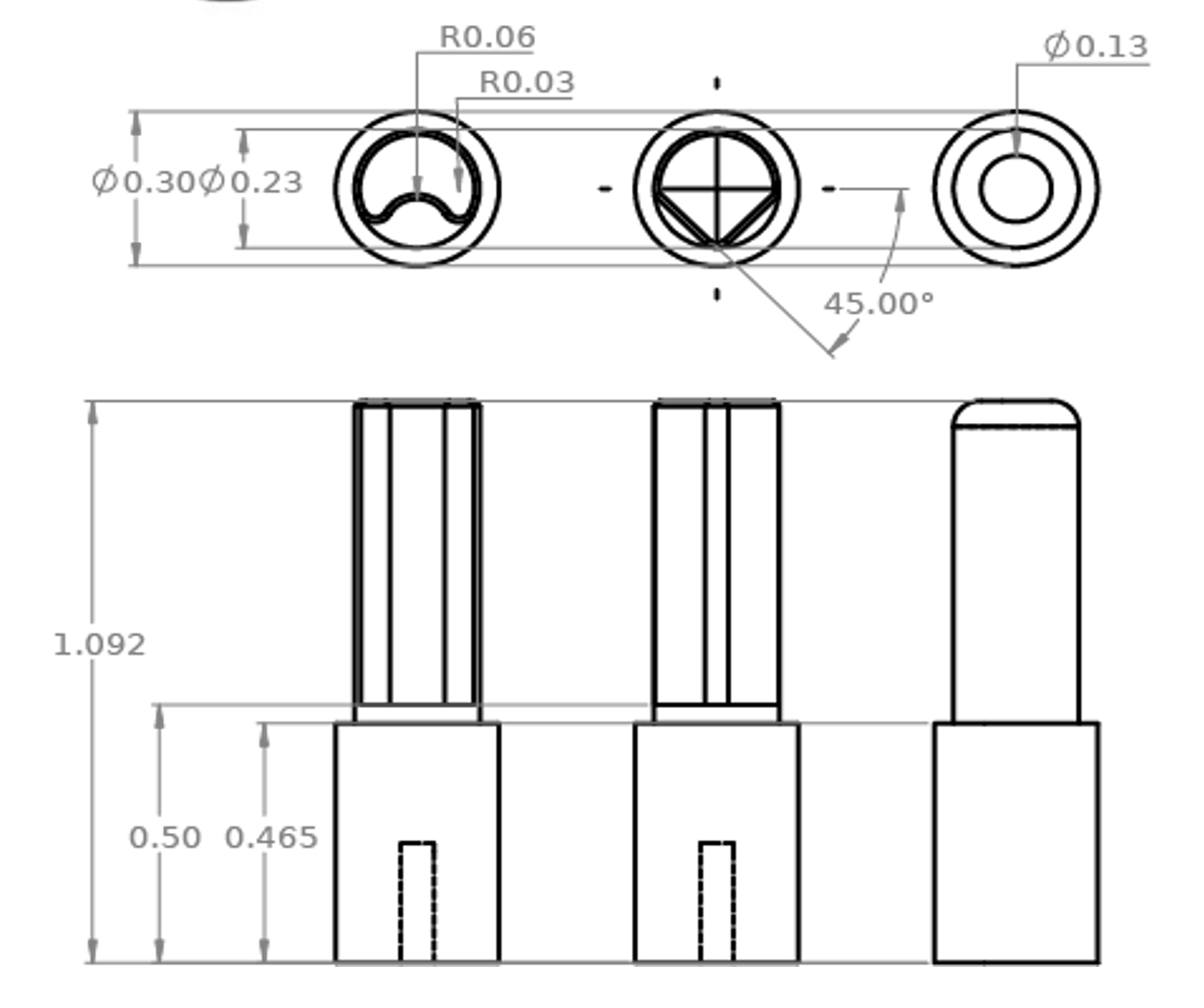}
        \caption{Technical drawings and dimensions of the concave, wedge-shaped, and cylindrical rod cross sections. These dimensions apply to the rods shown in Figs.~\ref{fig:rods_before_after}, \ref{fig:mds_imaging}(b), and \ref{fig:profiler}(a).}
    \end{subfigure}
    \caption{DiMES hardware used for stationary carbon rod ablation experiments.}
    \label{fig:rods_geometry}
\end{figure}

The first approach for measuring carbon ablation under high heat fluxes in the DIII-D tokamak focused on stationary carbon rods. For these measurements, the standard DiMES head was modified to hold up to three rods oriented approximately perpendicular to the plasma flow and arranged so that no rod directly shadowed another, as shown in Fig.~\ref{fig:rods_geometry}. Each rod was 1.5~cm in height and protruded into the DIII-D scrape-off layer. The rods were manufactured from ATJ graphite, and selected cylindrical rods were coated with SiC.

Three principal cross-sectional geometries were tested: a cylindrical rod representing a blunt body, a wedge-shaped rod resembling the Galileo probe forebody, and a concave rod with a cavity on the plasma-facing side. The concave geometry was included to examine whether local plasma detachment and geometric shadowing could reduce the incident heat flow. Some heads also used reversed (``backward'') orientations of the wedge or concave rods. Five DiMES-head configurations were used during the experimental campaign, as summarized in Table~\ref{tab:rod_configurations}. A more detailed description of the mechanical design is given in \cite{Orlov2021IMECE}.

\begin{table*}[htbp]
    \centering
    \caption{Rod configurations used during the stationary-ablation campaign. Slot numbers follow the DiMES-head nomenclature shown in Fig.~\ref{fig:rods_geometry}. ``Backward'' denotes installation of an asymmetric rod in the direction opposite to its nominal plasma-facing orientation.}
    \label{tab:rod_configurations}
    \begin{tabular}{cclll}
        \hline\hline
        Head & DIII-D discharges & Slot 1 & Slot 2 & Slot 3 \\
        \hline
        1 & 186095, 186096 & Backward wedge & Cylindrical & Backward concave \\
        2 & 186097, 186098 & SiC-coated cylindrical & Cylindrical & SiC-coated cylindrical \\
        3 & 186156, 186157 & Concave & Cylindrical & Wedge \\
        4 & 186158         & Wedge & Cylindrical & Cylindrical \\
        5 & 186159, 186160 & SiC-coated cylindrical & Cylindrical & SiC-coated cylindrical \\
        \hline\hline
    \end{tabular}
\end{table*}

\begin{figure}[htbp]
\centering
    \begin{subfigure}[b]{0.47\textwidth}
        \centering
        \includegraphics[width=\textwidth]{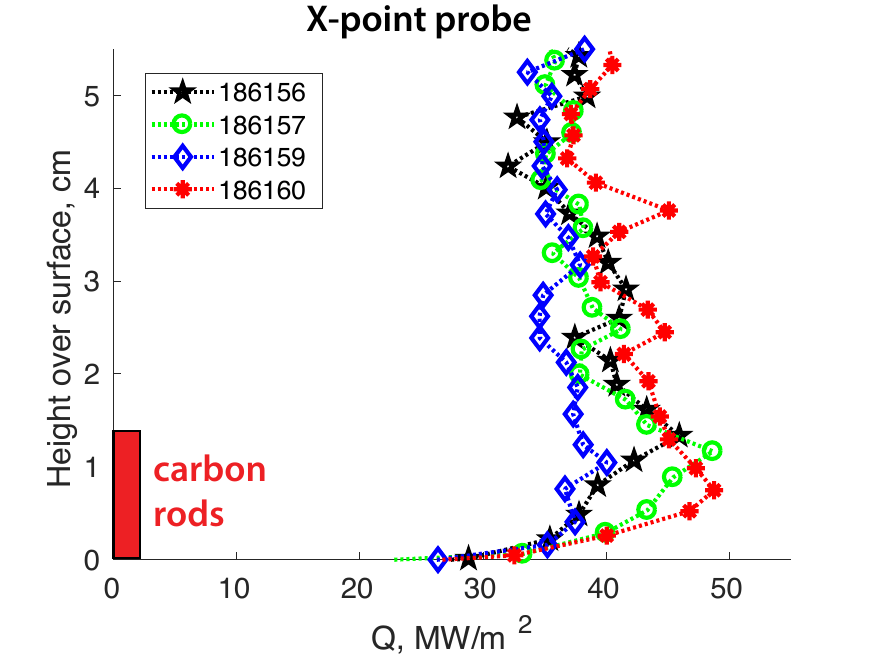}
        \caption{Heat flux profile from the X-point reciprocating Langmuir probe.}
    \end{subfigure}
    \hfill
    \begin{subfigure}[b]{0.47\textwidth}
        \centering
        \includegraphics[width=\textwidth]{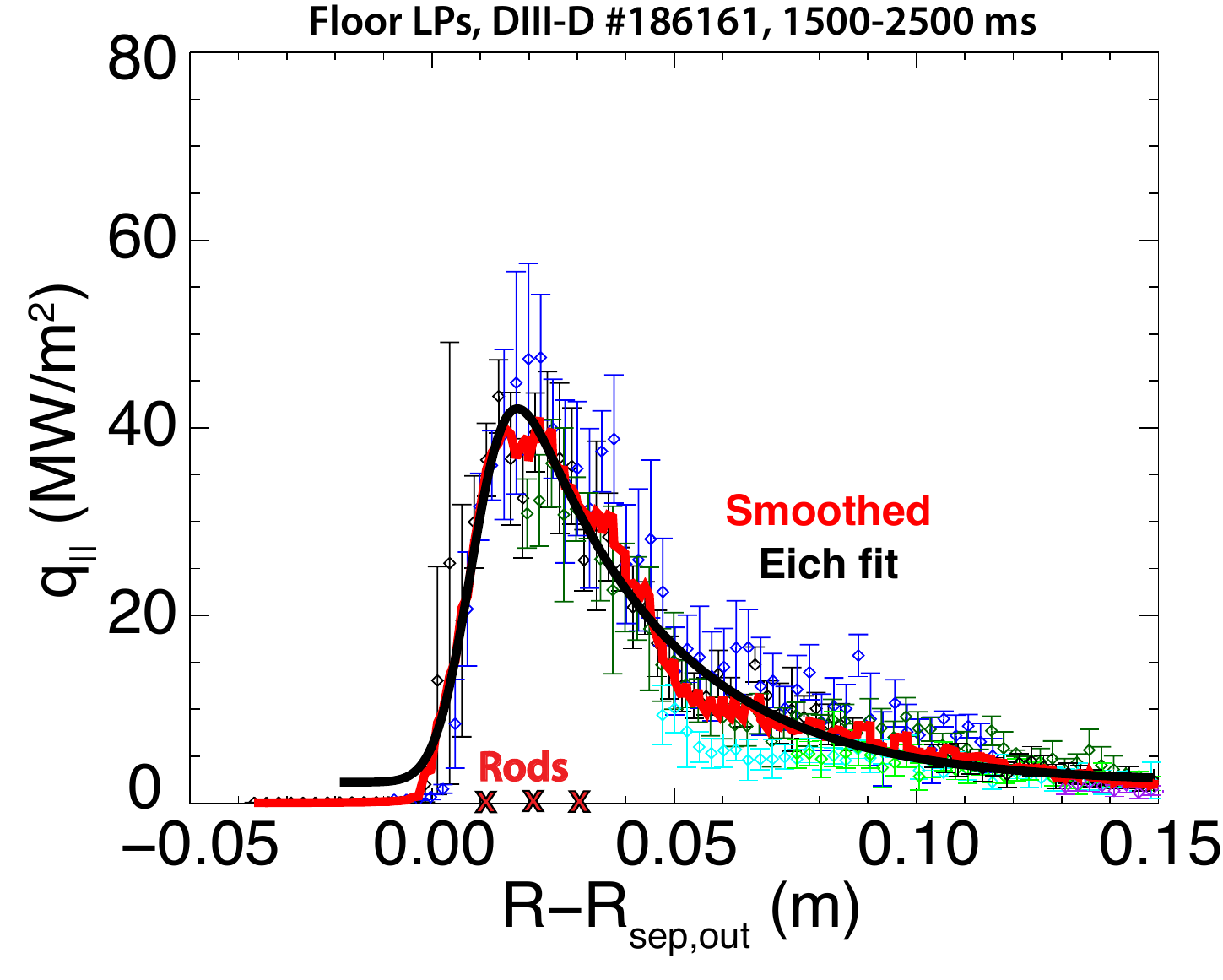}
        \caption{Mapped divertor heat flux profile and Eich fit during the strike-point sweep.}
    \end{subfigure}
    \caption{Measured heat fluxes in the divertor region near the DiMES location prior to the carbon rod ablation experiments.}
    \label{fig:heatflux}
\end{figure}

Experiments were conducted in lower single-null (LSN) discharges in L-mode, with the outer strike point (OSP) positioned near the DiMES port on the lower divertor shelf. Prior to the ablation experiments, the plasma conditions in the divertor region were characterized using the x-point reciprocating probe and the fixed floor Langmuir probes. The reciprocating probe measured the local plasma density, temperature, and flow velocity as a function of height above the divertor surface. Reciprocating Langmuir probe data, illustrated in Fig.~\ref{fig:heatflux}(a), indicated a parallel heat flux of approximately 40~MW\,m$^{-2}$ at 1.5~cm above the surface, with similar conditions observed across multiple discharges.

To relate these measurements to the rod locations, a strike-point sweep was performed and mapped onto a stationary equilibrium. The resulting heat flux profile across the divertor floor, obtained from the fixed floor Langmuir Probes, together with an Eich fit \cite{Eich2013}, is shown in Fig.~\ref{fig:heatflux}(b). The Eich profile characterization fits the data by convolving an exponential decay of parallel heat flux in the scrape-off layer with a Gaussian smoothing function that accounts for cross-field heat diffusion into the private flux region. The maximum heat flux occurred slightly radially outward of the nominal OSP, coinciding with the positions of the carbon rods. This analysis confirmed that the rods experienced parallel heat fluxes in the range of 30--40~MW\,m$^{-2}$ during the stationary phases of the discharge. During the rod exposure experiments, the outer strike point was positioned adjacent to the DiMES location and held approximately stationary throughout the plasma flattop phase. This stationary phase lasted about 4.5 s, extending from the onset of flattop conditions until the beginning of plasma rampdown, thereby providing a nearly constant heat-flux environment for the ablation measurements.

The rod-ablation analysis combines divertor plasma diagnostics, post-exposure analysis, spectroscopy, and thermal modeling. DTS, reciprocating-probe, and floor-probe measurements were used to reconstruct the plasma conditions and heat flux incident on the rods. Material erosion was quantified using optical profilometry and pre-/post-exposure mass measurements, while MDS and visible imaging constrained the timing of active carbon sublimation. ANSYS finite-element calculations were used to estimate rod surface temperatures under the measured heat loads. Together, these measurements provide the basis for the experimental mass-loss estimates and subsequent comparison with semi-empirical ablation models.


Most rod configurations were exposed in two consecutive plasma discharges, each having an approximately 4.5~s stationary phase and therefore providing about 9~s of cumulative high-heat-flux exposure. Head~4 was exposed only in discharge~186158. The remainder of this discussion first focuses on Head~5, which contained three cylindrical rods and was exposed in discharges~186159 and 186160. The rods in slots~1 and 3 were SiC-coated, whereas the middle rod in slot~2 was uncoated ATJ graphite.

\begin{figure}[htbp]
    \centering
    \begin{subfigure}[b]{0.4\textwidth}
        \centering
        \includegraphics[width=\textwidth]{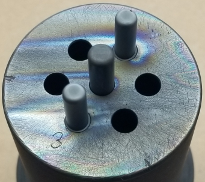}
        \caption{Head~5 before exposure. The cylindrical rods in slots~1 and 3 were SiC-coated, whereas the rod in slot~2 was uncoated ATJ graphite.}
    \end{subfigure}
    \hfill
    \begin{subfigure}[b]{0.4\textwidth}
        \centering
        \includegraphics[width=\textwidth]{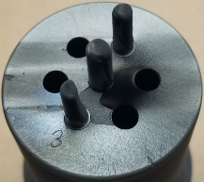}
        \caption{Head~5 after exposure to high-heat-flux plasma conditions in discharges~186159 and 186160.}
    \end{subfigure}
    \caption{Head~5 before and after the stationary carbon-rod ablation experiments. The slot-3 rod, located closest to the OSP, exhibited the greatest mass loss and visible recession. Dimensions of the cylindrical rods are provided in Fig.~\ref{fig:rods_geometry}(b).}    \label{fig:rods_before_after}
\end{figure}

Exposure produced immediate and visible ablation. Photographs of Head~5 before and after exposure are shown in Fig.~\ref{fig:rods_before_after}. The slot-3 rod, located closest to the OSP and appearing at the lower left in the figure, experienced the greatest visible recession. Its mass decreased by 0.1208~g, from 1.6551~g before exposure to 1.5343~g afterward.


\begin{figure}[htbp]
    \centering
    \begin{subfigure}[b]{0.35\textwidth}
        \centering
        \includegraphics[width=\textwidth]{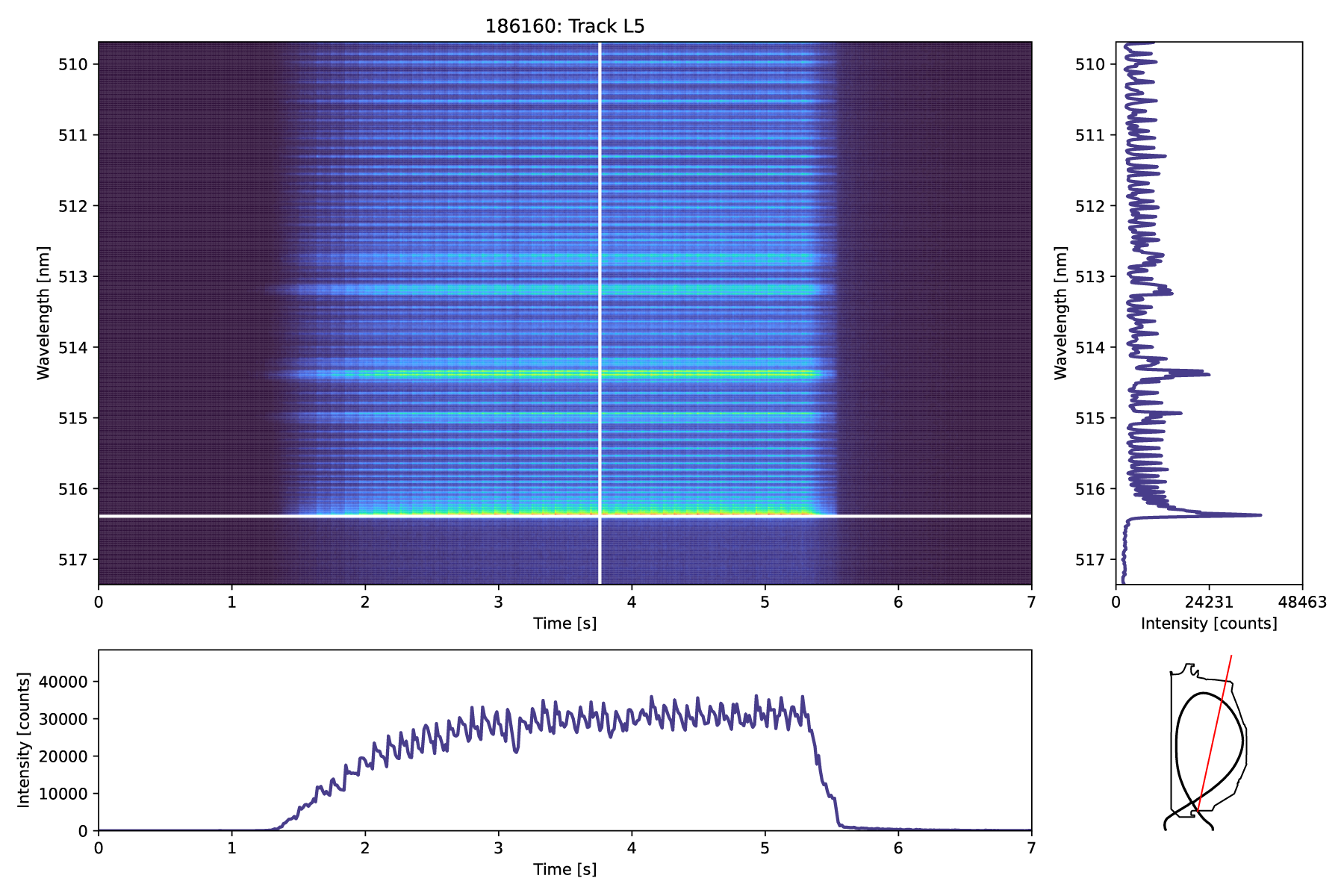}
        \caption{MDS spectrogram for a chord viewing the DiMES port in discharge 186160, showing strong C$_2$ Swan band emission.}
    \end{subfigure}
    \hfill
    \begin{subfigure}[b]{0.3\textwidth}
        \centering
        \includegraphics[width=\textwidth]{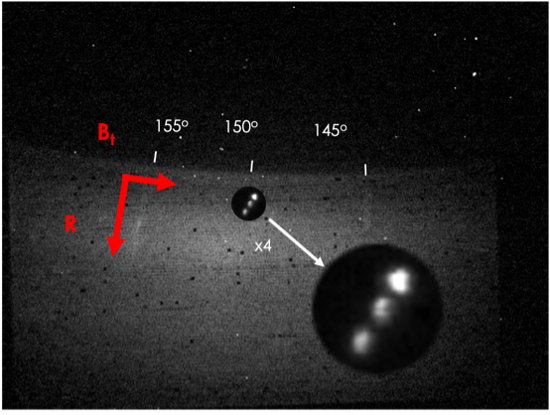}
        \caption{Visible fast-framing camera image of emission from the three ablation rods. Dimensions of the rods are provided in Fig.~\ref{fig:rods_geometry}(b).}
    \end{subfigure}
    \hfill
    \begin{subfigure}[b]{0.35\textwidth}
        \centering
        \includegraphics[width=\textwidth]{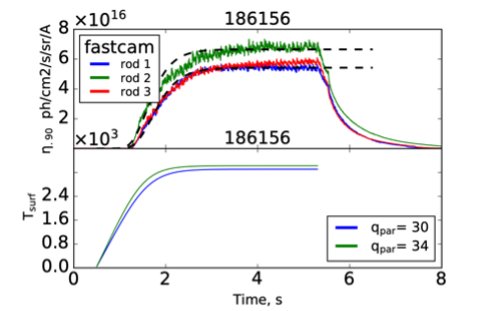}
        \caption{Time history of absolute narrow-band visible emission (top, in spectral radiance units of $\text{photons}\cdot\text{cm}^{-2}\cdot\text{s}^{-1}\cdot\text{sr}^{-1}\cdot\text{\AA}^{-1}$) and calculated history of inferred rod surface temperature (bottom, in $\times 10^3$~K), showing the experimental surface temperature cleanly stabilizing within the expected graphite sublimation range of approximately 3{,}700--3{,}800~K.}
    \end{subfigure}
    \caption{Spectroscopic and imaging diagnostics of carbon rod ablation at the DiMES port.}
    \label{fig:mds_imaging}
\end{figure}

Carbon ablation was further confirmed by the Multichordal Divertor Spectrometer (MDS) using a chord that directly views the DiMES port. In discharge~186160, strong carbon emission appeared around 2~s, as indicated by molecular emission in the 510--517~nm range, see Fig.~\ref{fig:mds_imaging}(a). The emission signal remained elevated throughout the ablation phase, providing an independent measure of the ablation duration. Spectral analysis during the flattop phase revealed strong bands around 516--520~nm, corresponding to the C$_2$ Swan bands \cite{Herzberg1950}, which indicate rapid sublimation of carbon directly from the solid to the gas phase. Similar emission features have been reported in plasma wind-tunnel studies of ablators \cite{Wernitz2011AIAA} and in previous DIII-D experiments \cite{McLean2009PhD}.

In addition to spectroscopy, a visible fast-framing camera with a vertical view of the DiMES port recorded the ablation process, as shown in Fig.~\ref{fig:mds_imaging}(b). The three bright spots near the toroidal angle of 150$^\circ$ correspond to the ablation rods. The time history of the integrated brightness from the camera and the inferred surface temperature evolution are shown in Fig.~\ref{fig:mds_imaging}(c). The surface temperature is estimated from narrow-band visible emission centered at 601.5~nm, which is dominated by continuum thermal radiation from the hot graphite surface, with a comparatively small contribution from nearby carbon line emission. Under the assumption that the incident parallel heat flux remains approximately constant during the stationary phase of the discharge, the temporal evolution of the measured brightness is interpreted using a grey-body radiation model with an effective emissivity representative of graphite.

The rapid rise in the emission signal during the first $\sim$2~s corresponds to heating of the rod surface toward radiative equilibrium, while the subsequent saturation of the brightness indicates that a quasi-steady surface temperature has been reached. The inferred equilibrium temperature is consistent with steady-state energy balance for parallel heat fluxes in the range 30--34~MW\,m$^{-2}$, in agreement with independent heat-flux estimates obtained from Langmuir probe measurements. While this approach does not provide an absolute pyrometric temperature measurement, it robustly identifies the time required to reach thermal steady state and constrains the surface temperature evolution within the experimental uncertainty. Specifically, the inferred steady-state surface temperatures for the exposed graphite rods level off between approximately 3{,}700~K and 3{,}800~K, aligning closely with the material's expected sublimation threshold under these heat loads.

\begin{figure}[htbp]
    \centering
    \begin{subfigure}[b]{0.20\textwidth}
        \centering
        \includegraphics[width=\textwidth]{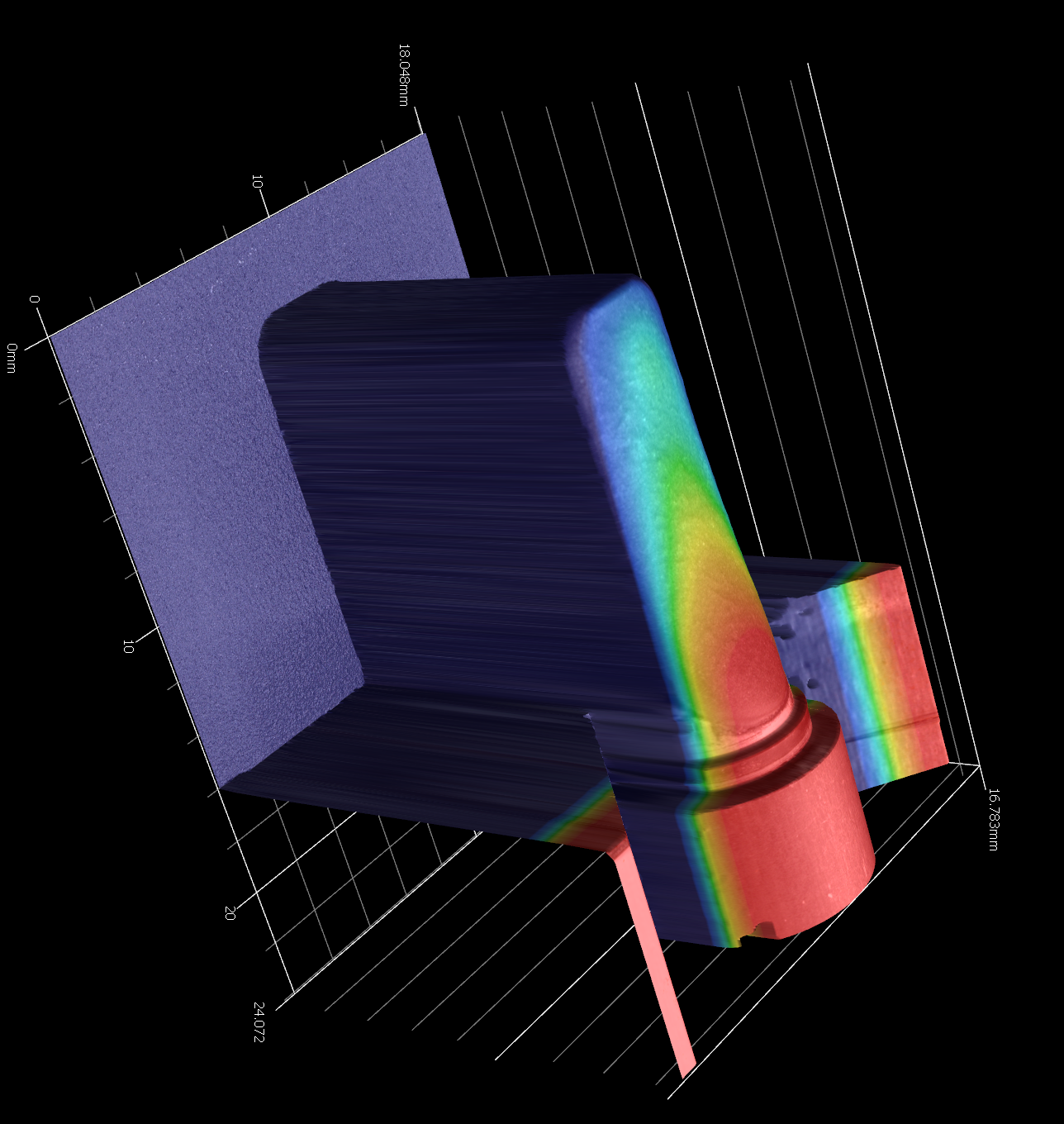}
        \caption{Three-dimensional reconstruction of the cylindrical rod surface from the optical profiler; color indicates height above a reference plane. The original rod dimensions are provided in Fig.~\ref{fig:rods_geometry}(b).}    \end{subfigure}
    \hfill
    \begin{subfigure}[b]{0.30\textwidth}
        \centering
        \includegraphics[width=\textwidth]{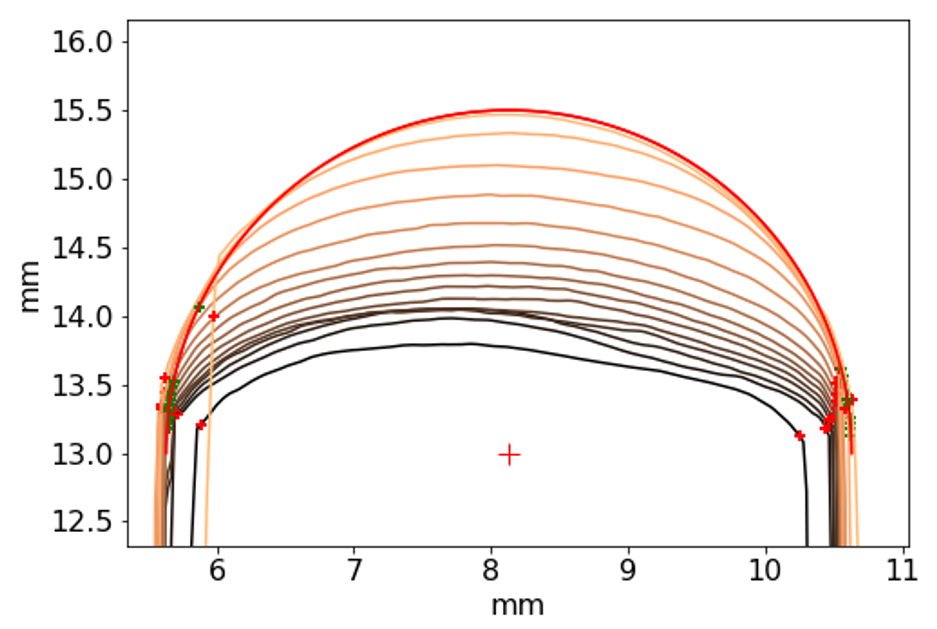}
        \caption{Cross-sectional profiles at different heights; bold orange curve is the original cylindrical profile.}
    \end{subfigure}
    \hfill
    \begin{subfigure}[b]{0.30\textwidth}
        \centering
        \includegraphics[width=\textwidth]{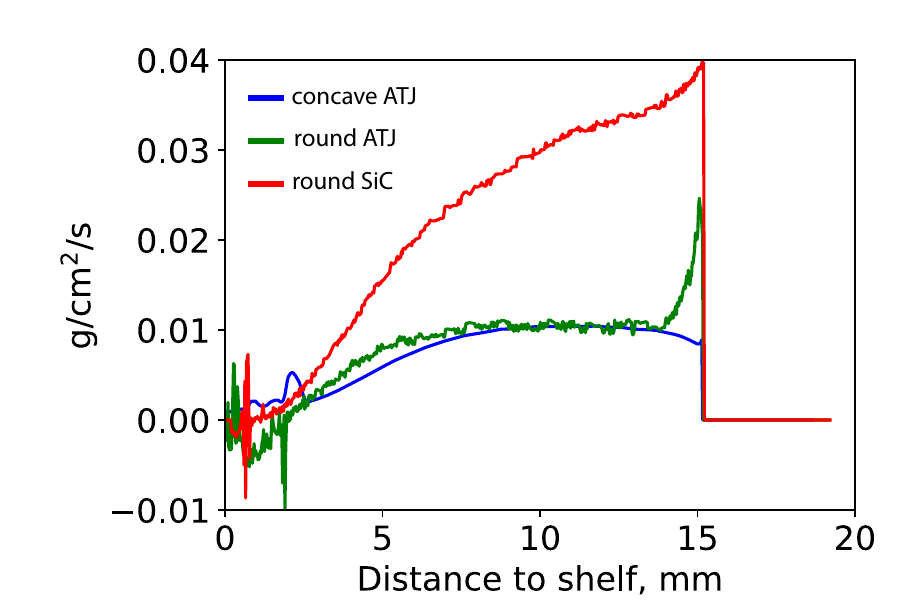}
        \caption{Average mass loss rate as a function of distance from the shelf for the three rod configurations: concave ATJ graphite rod, outboard slot (blue), round ATJ graphite rod, middle slot (green), and round SiC-coated ATJ graphite rod, innermost slot (red).}
    \end{subfigure}
    \caption{Post-mortem characterization of plasma-exposed carbon rods using an optical 3D surface profiler. The measurements are used to reconstruct the rod geometry, quantify material recession, and determine the local mass loss rate.}
    \label{fig:profiler}
\end{figure}

After the DIII-D experiments, the rods were removed from the DiMES head for detailed post-mortem analysis. Surface recession profiles were measured using an optical 3D surface profiler. A reconstructed 3D surface of one of the cylindrical rods is shown in Fig.~\ref{fig:profiler}(a), where color indicates the height above a reference plane. Regions with the smallest height correspond to locations of the largest surface recession and highest local heat flux. Cross-sectional profiles taken at different heights along the rod axis are shown in Fig.~\ref{fig:profiler}(b). The bold orange line denotes the original pre-exposure cylindrical profile, while darker profiles indicate greater ablation near the rod top.

From the measured recession patterns and the known exposure time, the average surface mass loss rate $\dot{m}''$ (in g\,cm$^{-2}$\,s$^{-1}$) was estimated as
\begin{equation}
    \dot{m} = \frac{\Delta m}{A\,\Delta t},
    \label{eq:massloss_rod}
\end{equation}
where $\Delta m$ is the measured mass loss, $A$ is the initial surface area of the exposed rod, and $\Delta t$ is the total exposure time. 

The resulting mass loss rates for the three rod shapes are summarized in Fig.~\ref{fig:profiler}(c). The wedge-shaped rod exhibited the largest mass loss rate, $(1$--$3)\times10^{-2}$~g\,cm$^{-2}$\,s$^{-1}$, while the cylindrical and concave rods showed similar lower values of $(1.03$--$1.06)\times10^{-2}$~g\,cm$^{-2}$\,s$^{-1}$. These experimentally inferred mass loss rates are comparable to the rates predicted by the Park and Matsuyama ablation models discussed in Sec.~\ref{sec:ablation_models}. For reference, the Matsuyama model predicts mass loss rates of $(0.66$--$0.87)\times10^{-2}$~g\,cm$^{-2}$\,s$^{-1}$, while the Park model predicts $(1.28$--$1.70)\times10^{-2}$~g\,cm$^{-2}$\,s$^{-1}$ under the measured DIII-D heat-flux conditions. The experimentally measured mass loss rates are in closer agreement with the Park-model predictions, whereas the Matsuyama model tends to underpredict the values. Quantitatively, the cylindrical and concave rods differ from the Park-model predictions by approximately 20--40\%, whereas the Matsuyama model underpredicts the measured ablation rates by roughly 20--50\%. While the measured ablation rate of the wedge-shaped rod partially brackets the upper bounds of the Park model, its peak enhanced ablation completely outpaces both model ranges. This demonstrates that the Park model provides a far more accurate baseline approximation for this geometry, though local heat-flux concentration effects from the sharp leading edges still introduce an enhanced recession rate not explicitly captured by either standard model.

The enhanced ablation of the wedge-shaped rod is likely associated with geometry-dependent heat-flux concentration effects that are not explicitly captured by either semi-empirical model. In particular, the sharp leading edges and reduced local radius of curvature of the wedge geometry can increase local heat loading and produce stronger surface-temperature gradients, resulting in enhanced recession in the most strongly exposed regions. In contrast, the cylindrical and concave geometries distribute the incident heat flux more uniformly over the exposed surface and exhibit similar average mass loss rates. These initial results suggest that Park's model provides a reasonable baseline approximation for the blunt and concave geometries, but still fails to capture the significantly enhanced mass loss rates exhibited by the wedge geometry. Conversely, the Matsuyama model systematically underpredicts the ablation values across all tested target profiles, highlighting the necessity of integrating geometry-dependent local heat-flux concentration effects into future material validation frameworks. 

\section{Ablation of carbon pellets in plasma core region}
\label{sec:pellets}

\subsection{Experimental setup}

A second approach to studying carbon ablation under extreme heat fluxes in DIII-D employed slowly moving graphite pellets launched from the DiMES port and tracked as they traversed the edge and core plasma. In contrast to the stationary rods discussed in Sec.~\ref{sec:rods}, this method follows the evolution of an individual ablator element along its full trajectory through the scrape-off layer and across the separatrix, observing its complete ablation within the field of view of the diagnostics.

\begin{figure}[htbp]
    \centering
    \begin{subfigure}[b]{0.3\textwidth}
        \centering
        \includegraphics[width=\textwidth]{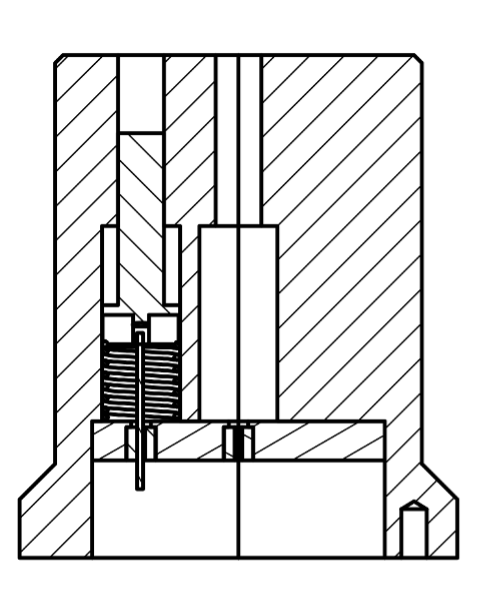}
        \caption{Schematic of the DiMES pellet launcher showing one of the shafts with a piston and a precompressed spring.}
    \end{subfigure}
    \hfill
    \begin{subfigure}[b]{0.3\textwidth}
        \centering
        \includegraphics[width=\textwidth]{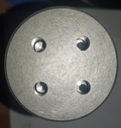}
        \caption{Top-down view of the DiMES pellet launcher installed behind the DiMES port, showing the four pellet launch channels.}
    \end{subfigure}
    \caption{Experimental setup for the carbon-pellet ablation studies in DIII-D. The circular DiMES head has a standard diameter of approximately 5 cm, providing a dimensional reference for both panels.}
    \label{fig:pellet_setup}
\end{figure}

The pellets were launched using a dedicated DiMES pellet launcher installed in the DiMES port \cite{Orlov2021IMECE}. The launcher consisted of a linear guide and pusher mechanism that drove the pellet along a vertical tube aligned with the DiMES aperture. The actuation was based on an electrically triggered release mechanism in which a current pulse vaporized a restraining wire, freeing a spring-loaded piston that rapidly accelerated the pellet along the guide tube. Typical pellet velocities at the port exit were in the range of 3--10~m\,s$^{-1}$, as determined from time-resolved imaging of the pellet trajectory. The launcher was operated remotely from the DIII-D control room, with mechanical reloading performed between discharges without additional vacuum breaks. A schematic of the launcher geometry, together with a photograph of the installed hardware, is shown in Fig.~\ref{fig:pellet_setup}. The launcher assembly was installed within a standard DiMES head, which is approximately 5 cm in diameter.

In the absence of plasma flow and magnetic forces, a pellet launched vertically from the DiMES port would travel along a straight line determined purely by gravity and the mechanical alignment of the launcher. In the presence of the DIII-D edge plasma, however, the pellet experiences a combination of forces due to gravity, ion drag due to plasma flows parallel to the magnetic field, $\mathbf{E}\times\mathbf{B}$ drifts, and $\mathbf{j}\times\mathbf{B}$ force induced by charged plasma particles. The pellet is also heated and ablated by the plasma along its path. The resulting trajectory, therefore, provides a sensitive probe for both the plasma flow field and the ablation dynamics of the pellet.

Two types of graphite pellets were used in this work: a porous ``tidal'' graphite, representative of lower-density ablators ($\rho = 0.8\text{~g~cm}^{-3}$), and a dense glassy carbon ($\rho = 1.4\text{~g~cm}^{-3}$). Both types of pellets possessed spherical shapes; the compact glassy pellets had diameters of approximately $0.9\text{~mm}$, whereas the larger porous tidal pellets exhibited a wider diameter range of $1.8$--$3.6\text{~mm}$ corresponding to their varying masses ($2.7$--$20.3\text{~mg}$). 
Prior to the experiments, each pellet was weighed using a precision balance and its geometric dimensions were recorded in order to determine the initial mass and density. During the DIII-D experiments, a range of plasma conditions was tested, including variations in line-averaged density and heating power, providing a ``control knob'' for scanning the incident heat flux on the pellets and the strength of the plasma flows.

\subsection{Imaging diagnostics and observed trajectories}

\begin{figure}[htbp]
    \centering
    \begin{subfigure}[b]{0.47\textwidth}
        \centering
        \includegraphics[width=\textwidth]{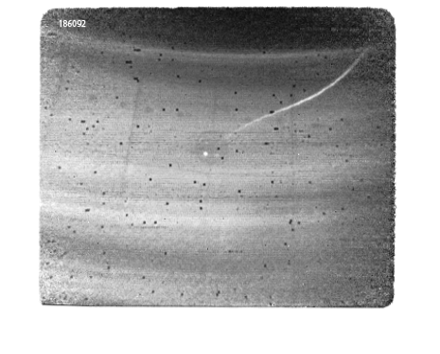}
        \caption{Visible fast-framing image of a porous carbon pellet launched from the DiMES port and moving toward the core plasma.}
    \end{subfigure}
    \hfill
    \begin{subfigure}[b]{0.4\textwidth}
        \centering
        \includegraphics[width=\textwidth]{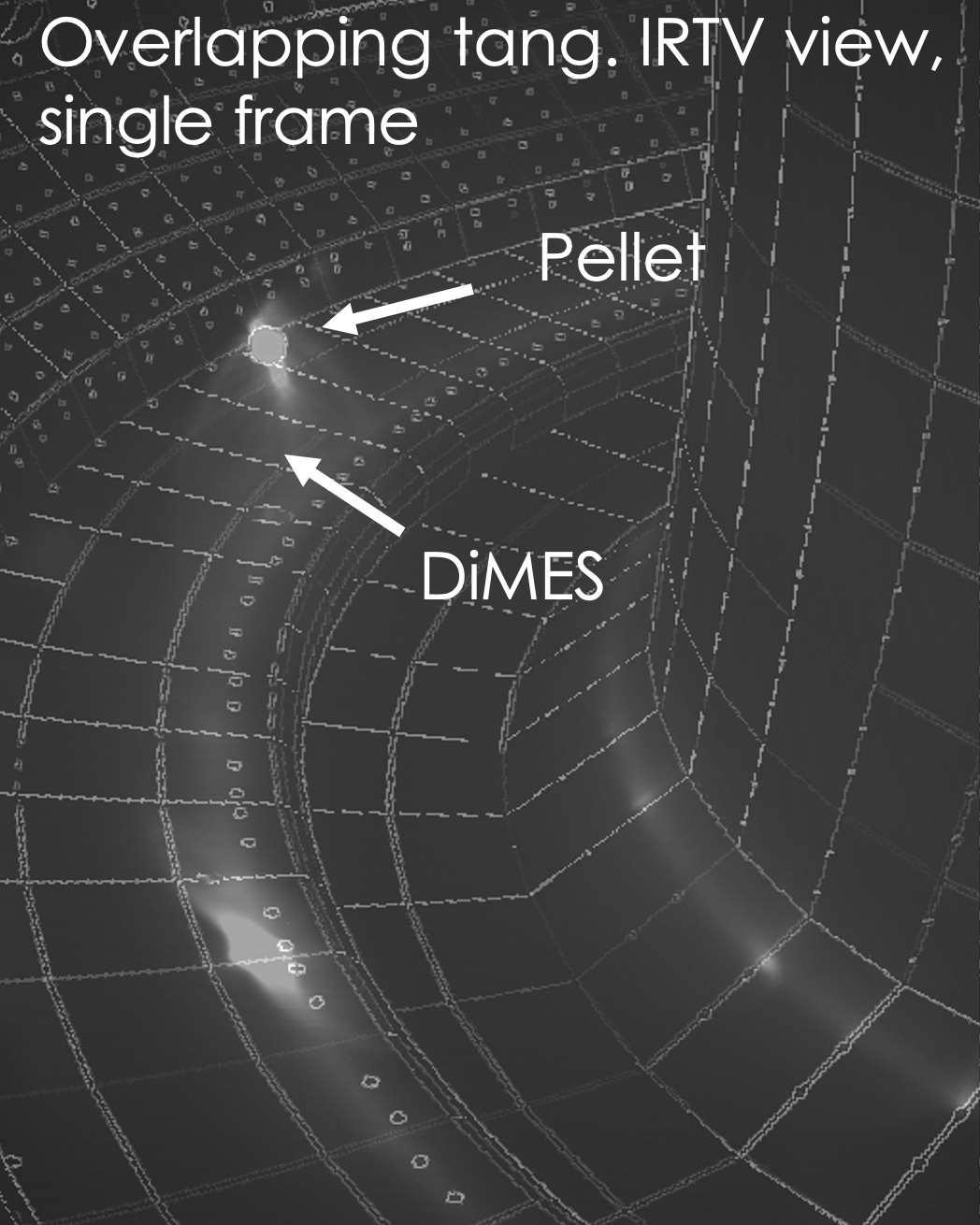}
        \caption{Infrared camera snapshot of the same pellet, showing thermal emission as it crosses the separatrix.}
    \end{subfigure}
    \caption{Imaging diagnostics for carbon pellet ablation experiments in discharge 186093.}
    \label{fig:pellet_traj}
\end{figure}

The trajectory of a typical porous graphite pellet with an initial mass of 20.3~mg launched from the DiMES port in discharge~186093 is shown in Fig.~\ref{fig:pellet_traj}. The visible fast-framing camera provides a tangential view of the pellet path as it leaves the divertor region, crosses the separatrix, and enters the hotter regions of the core plasma. The collisions with background plasma electrons cause the ablated material to rapidly ionize. Locally, this ionized material expands along the magnetic field lines, forming a localized, luminous plume immediately surrounding the moving pellet. Over the full duration of the ablation phase, the sequence of these frames traces out the macroscopic path of the pellet, appearing as the extended, bright trajectory track visible in the time-composite image in Fig.~\ref{fig:pellet_traj}(a)

An infrared (IR) camera viewing the same region recorded the thermal emission from the pellet and surrounding plasma, as illustrated in Fig.~\ref{fig:pellet_traj}(b). The IR data are used to estimate the pellet surface temperature as a function of time, under the assumption that the pellet emits approximately as a graybody with an effective emissivity characteristic of graphite \cite{SiegelHowell}. Combining the visible and IR imaging allows reconstruction of both the pellet trajectory and its thermal history during ablation.

\subsection{Integrated analysis of pellet dynamics and ablation}
\label{subsec:data_workflow}

A comprehensive assessment of pellet behavior in the DIII-D plasma relies on a tightly coupled set of diagnostic measurements and numerical models. Each diagnostic contributes specific plasma or pellet observables, which together form a consistent basis for reconstructing pellet trajectories, evaluating ablation rates, and validating predictive physics models.

\begin{figure}[htbp]
    \centering
    \includegraphics[width=0.33\textwidth]{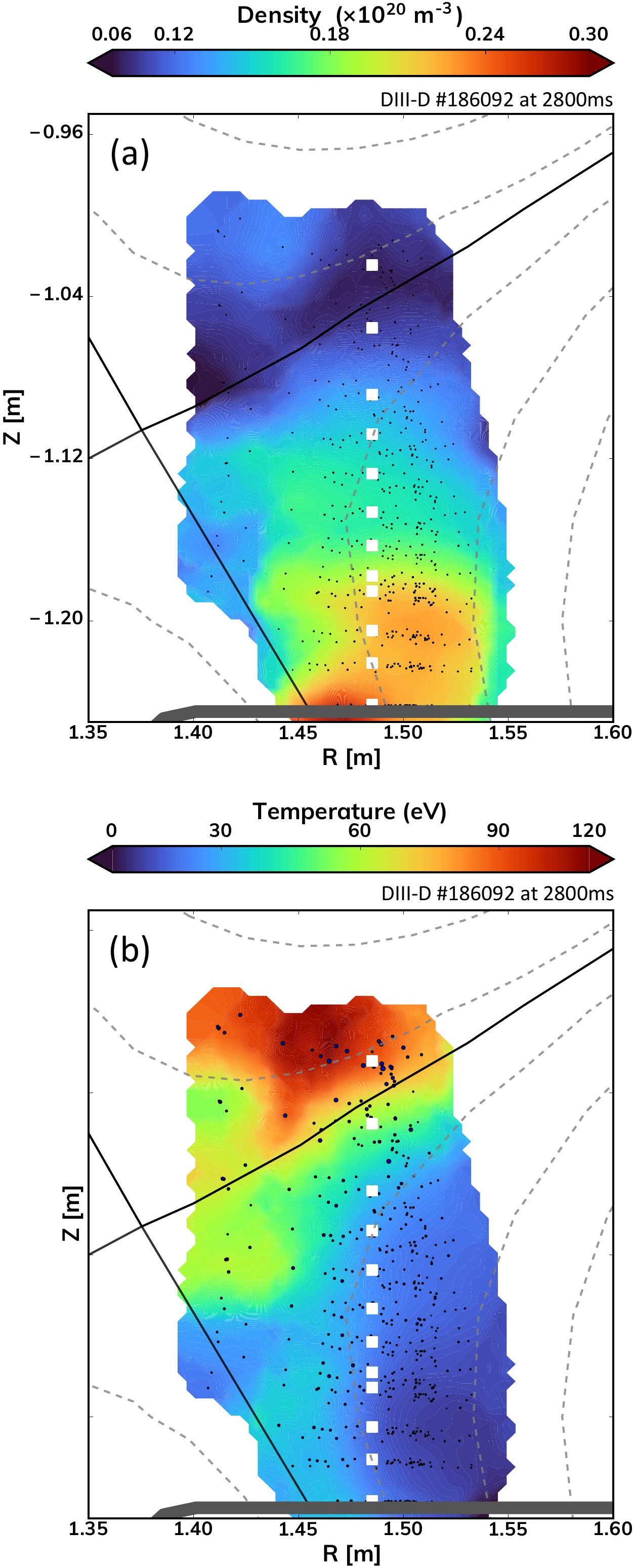}
    \caption{Divertor Thomson Scattering (DTS) maps for DIII-D discharge \#186092 showing (a) electron temperature and (b) electron density. The white squares indicate the locations of the 1D Divertor Thomson scattering measurement chords, while the black points represent the locations of data acquisition during the outer strike point (OSP) sweep. The DTS scans were performed during stationary flattop conditions and prior to pellet injection, providing plasma parameters for UEDGE modeling and constraints on divertor heat flux and ablation physics. At the 2800~ms timeslice, the OSP is located on the inboard side of the DiMES port.}    \label{fig:dts_map}
\end{figure}

Prior to pellet injection, the plasma environment near the DiMES port was characterized through controlled sweeping of the outer strike point (OSP) across the divertor region. These OSP sweeps were performed once the discharge reached steady flattop conditions and lasted approximately $1$~s, ensuring high spatial coverage of the region where pellets subsequently traveled. Divertor Thomson Scattering (DTS) measurements obtained during these sweeps provide two-dimensional maps of electron temperature $T_e$ and density $n_e$ near the strike point. These maps characterize the stationary divertor plasma into which the pellets were injected and serve as a critical basis for both heat-flux evaluation and ablation model inputs, as illustrated in Fig.~\ref{fig:dts_map}.

Complementary measurements were supplied by the X-point reciprocating probe (XPP), which provided localized profiles of $T_e$, $n_e$, and parallel ion flow during brief plunges into the plasma boundary, and by divertor floor Langmuir probes, which measured the parallel heat flux $q_{\parallel}$ near the DiMES location. Collectively, these diagnostics constrain the UEDGE fluid simulations of the scrape-off layer and divertor plasma \cite{Rognlien1992}, establishing a consistent background plasma state for assessing pellet--plasma interactions.

The UEDGE plasma solution supplies spatial profiles of temperature, density, flow, and magnetic geometry that are used by the DUSTT dust/particle transport code to compute pellet motion and ablation in three dimensions. DUSTT is a 3D Monte-Carlo code that advances an ensemble of particles in a prescribed plasma background while self-consistently solving for their motion, charging, energy balance, and mass loss \cite{Smirnov2007,Pigarov2005}. The equation of motion includes a broad set of forces acting on a pellet: gravity, ion-drag (both collection and Coulomb components), neutral drag, electrostatic force from background electric fields, and the $\mathbf{j}\times\mathbf{B}$ force associated with the currents flowing through conducting pellet \cite{Smirnov2007}. For the millimeter-scale pellets considered here, gravity and ion-drag are the dominant forces, while the $\mathbf{j}\times\mathbf{B}$ term becomes important once the plasma currents collected by the pellet are substantially asymmetric due to plasma non-uniformity.

The thermal state and mass of the pellet evolve according to the dust-heating and ablation model originally developed for carbon grains in fusion edge plasmas \cite{Smirnov2007}. Heating terms include energy transfer from impinging plasma ions and electrons, surface recombination, and plasma radiation. Cooling terms include sublimation/vaporization, thermal radiation, and electron emission. 
The pellet  mass can change as a result of physical and chemical sputtering by impinging plasma ions, thermal sublimation/vaporization, and the deposition of background plasma impurities onto the pellet surface. These competing mass-loss and mass-gain processes define the rate of dust ablation in the $(n_e,T_e)$ parameter space. The DIII-D experiments described here operate firmly in the fast-sublimation regime, so that the pellets rapidly lose mass once they penetrate into the hot core.

At sufficiently high surface temperatures and ablation rates, a dense ablation cloud forms around the pellet. In this regime, simple orbital-motion-limited (OML) heat-flux estimates are no longer adequate, and shielding of the pellet surface by the cloud must be taken into account \cite{Krasheninnikov2009,Brown2014,Krasheninnikov2014,Marenkov2014}. In the present work this effect is implemented in DUSTT for carbon pellets in an ad-hoc manner (Sec.~\ref{subsec:pellet_imaging_modeling}) and used to reproduce experimentally inferred ablation rates.

In the fully coupled DUSTT--UEDGE framework, impurity sources from pellet ablation feed back on the plasma state, and the plasma solution is advanced in time with dust sources included \cite{Smirnov2020}. For the present analysis, however, the pellets are treated as test particles that do not significantly perturb the global plasma profiles, so that a one-way coupling is sufficient: UEDGE provides a stationary background, and DUSTT uses that background to evolve pellet trajectories and mass loss. This approach still retains the full 3D physics of pellet motion and ablation while allowing a direct, shot-by-shot comparison between measured and modeled pellet behavior.


\subsection{Comparison between UEDGE--DUSTT modeling and experimental ablation estimates}
\label{subsec:pellet_imaging_modeling}

The injection and subsequent ablation of the millimeter-size carbon pellets are directly observed using fast visible-light imaging and a tangential-view infrared (IR) camera. The visible fast-framing camera provides a top-down view of the pellet trajectory with high spatial resolution, enabling a clear tracking of its path projected onto the poloidal plane, as shown previously in Fig.~\ref{fig:pellet_traj}(a). Meanwhile, a tangential IR camera simultaneously images the pellet from the outboard side, recording its thermal radiation as the pellet heats in the plasma and generating a complementary trajectory projection in the $(R,Z)$ plane.

\begin{figure}[htbp]
    \centering
    \includegraphics[width=0.45\textwidth]{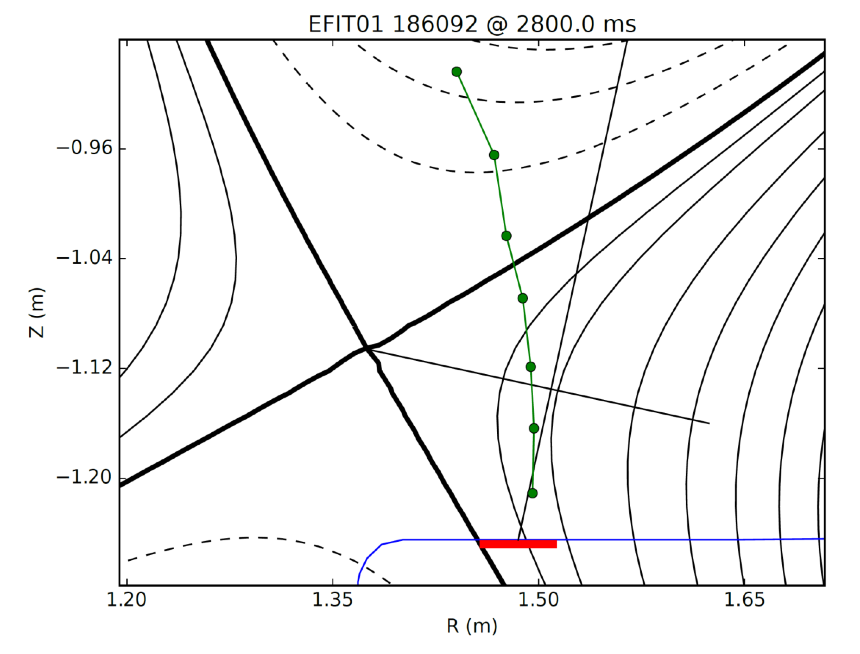}
    \caption{Experimentally reconstructed poloidal projection of a carbon pellet trajectory in discharge~186093 obtained using stereoscopic imaging from the visible fast-framing camera and tangential IR camera. The full trajectory is reconstructed in three dimensions using calibrated camera geometries and line-of-sight transformations, while only the $(R,Z)$ projection is shown here for clarity. Red line on this plot represents the location of DiMES, while the green dots are the successive locations of the pellet.}
\label{fig:pellet_3d}
\end{figure}

\begin{figure}[htbp]
    \centering
    \begin{subfigure}[b]{0.47\textwidth}
        \centering
        \includegraphics[width=\textwidth]{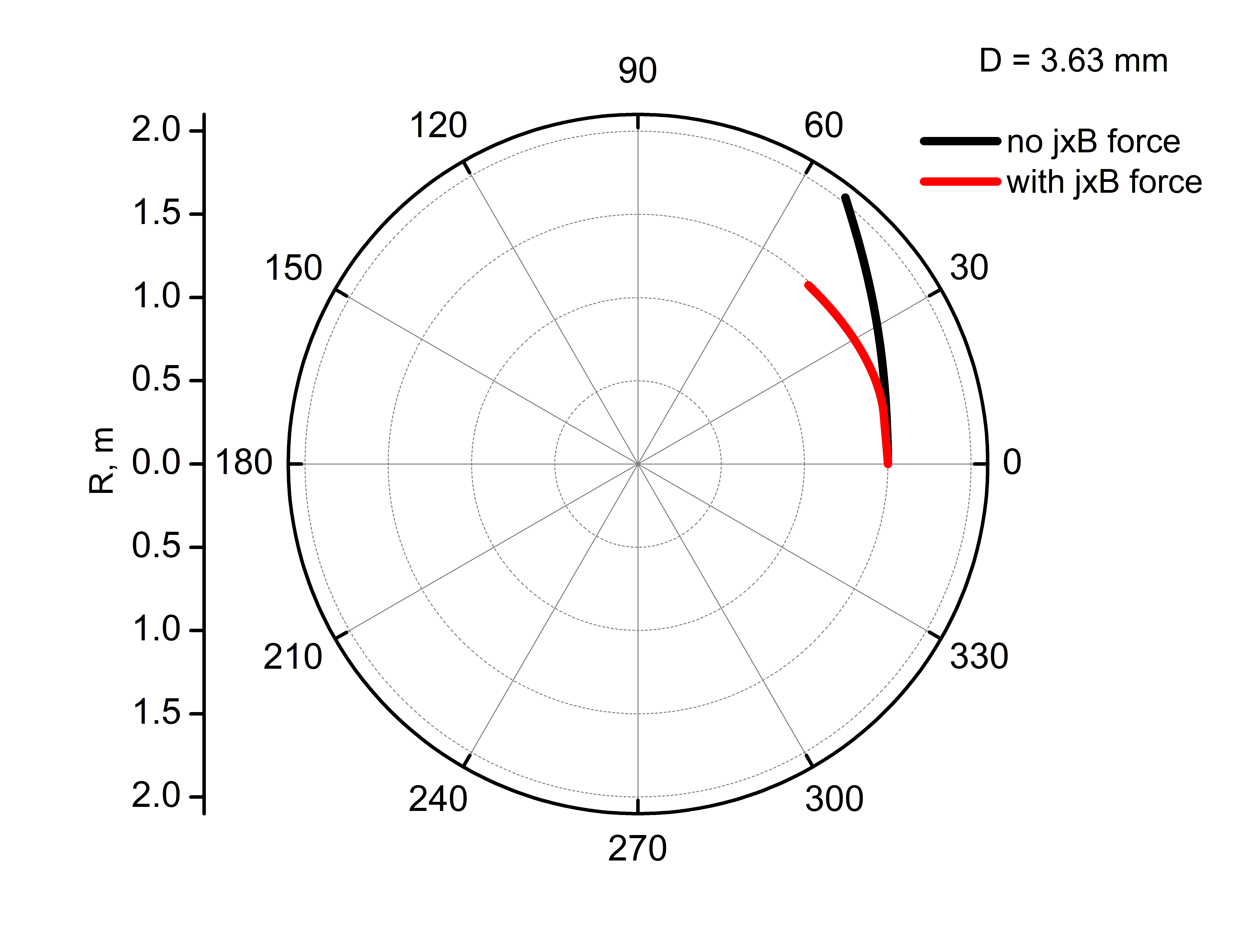}
        \caption{Top-view of UEDGE--DUSTT simulated trajectory for a 3.6~mm pellet in discharge~186092.}
        \label{fig:pellet_dustt_top}
    \end{subfigure}
    \hfill
    \begin{subfigure}[b]{0.47\textwidth}
        \centering
        \includegraphics[width=\textwidth]{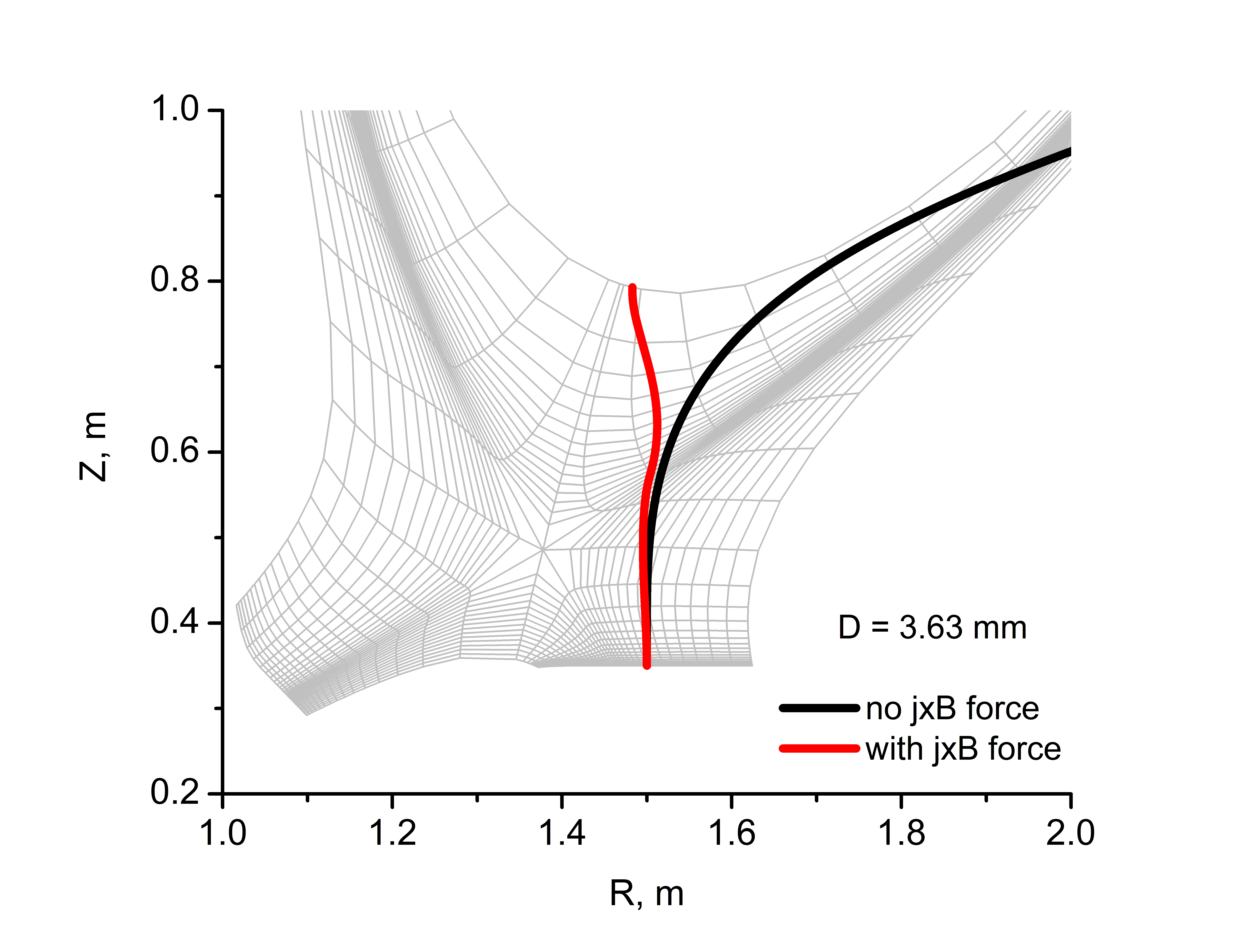}
        \caption{Side-view of UEDGE--DUSTT simulated trajectory for a 3.6~mm pellet in discharge~186092.}
        \label{fig:pellet_dustt_side}
    \end{subfigure}
    \caption{UEDGE--DUSTT simulated pellet trajectories for a 3.6~mm, 20.3~mg carbon pellet in discharge~186092 with and without the $\mathbf{j}\times\mathbf{B}$ force. Inclusion of $\mathbf{j}\times\mathbf{B}$ increases the predicted penetration depth and produces a trajectory more consistent with the experimentally observed behavior. The experimentally reconstructed trajectory is shown separately in Fig.~\ref{fig:pellet_3d}.}
    \label{fig:pellet_dustt_traj}
\end{figure}

To obtain the full three-dimensional trajectory, image centroids from both cameras are independently detected and then registered to a magnetic equilibrium geometry using measured camera calibrations and line-of-sight transformations. Interpolation between asynchronous camera frames is used when necessary to synchronize the pellet positions in time. This stereoscopic approach enables reconstruction of the pellet trajectory in $(R,Z,\phi)$ space. Figure~\ref{fig:pellet_3d} shows the experimentally reconstructed poloidal projection of the trajectory, demonstrating that the pellet crosses the separatrix and penetrates well into the confined plasma region before fully ablating. The corresponding three-dimensional trajectory information is subsequently used for comparison with the UEDGE--DUSTT modeling results presented in Fig.~\ref{fig:pellet_dustt_traj}.

In the present work, the pellet centroid was identified manually in each frame of the two cameras. However, the method is well suited for automation using modern image processing and tracking algorithms, which would enable statistical analysis over many shots and improved benchmarking of numerical models.

The reconstructed pellet trajectory provides a direct benchmark for the UEDGE--DUSTT pellet transport model. Figure~\ref{fig:pellet_dustt_traj} shows simulated trajectories for a 3.6~mm, 20.3~mg carbon pellet in discharge~186092. The top-view result in Fig.~\ref{fig:pellet_dustt_top} shows the toroidal and radial portion of the pellet trajectory driven by parallel plasma flow and pellet inertia. The poloidal-view trajectories in Fig.~\ref{fig:pellet_dustt_side} highlight the critical role of the $\mathbf{j}\times\mathbf{B}$ force acting on the conducting pellet. For micron-sized dust particles, this force is often negligible relative to other forces; however, for a millimeter-scale conducting pellet, the current induced by the plasma temperature gradient across it can be significant and the magnitude of $\mathbf{j}\times\mathbf{B}$ force becomes comparable to the radial centrifugal force. If this effect is omitted, the simulated pellet is rapidly lost to the wall and does not cross the separatrix, in stark contrast with the experiment. Inclusion of the $\mathbf{j}\times\mathbf{B}$ force redirects the pellet deeper into the plasma core and produces a trajectory that more closely resembles the experimentally reconstructed path shown in Fig.~\ref{fig:pellet_3d}. While a quantitative comparison is beyond the scope of the present work, the results suggest that electromagnetic forces contribute significantly to the observed pellet dynamics.


\begin{figure}[htbp]
    \centering
    \begin{subfigure}[b]{0.47\textwidth}
        \centering
        \includegraphics[width=\textwidth]{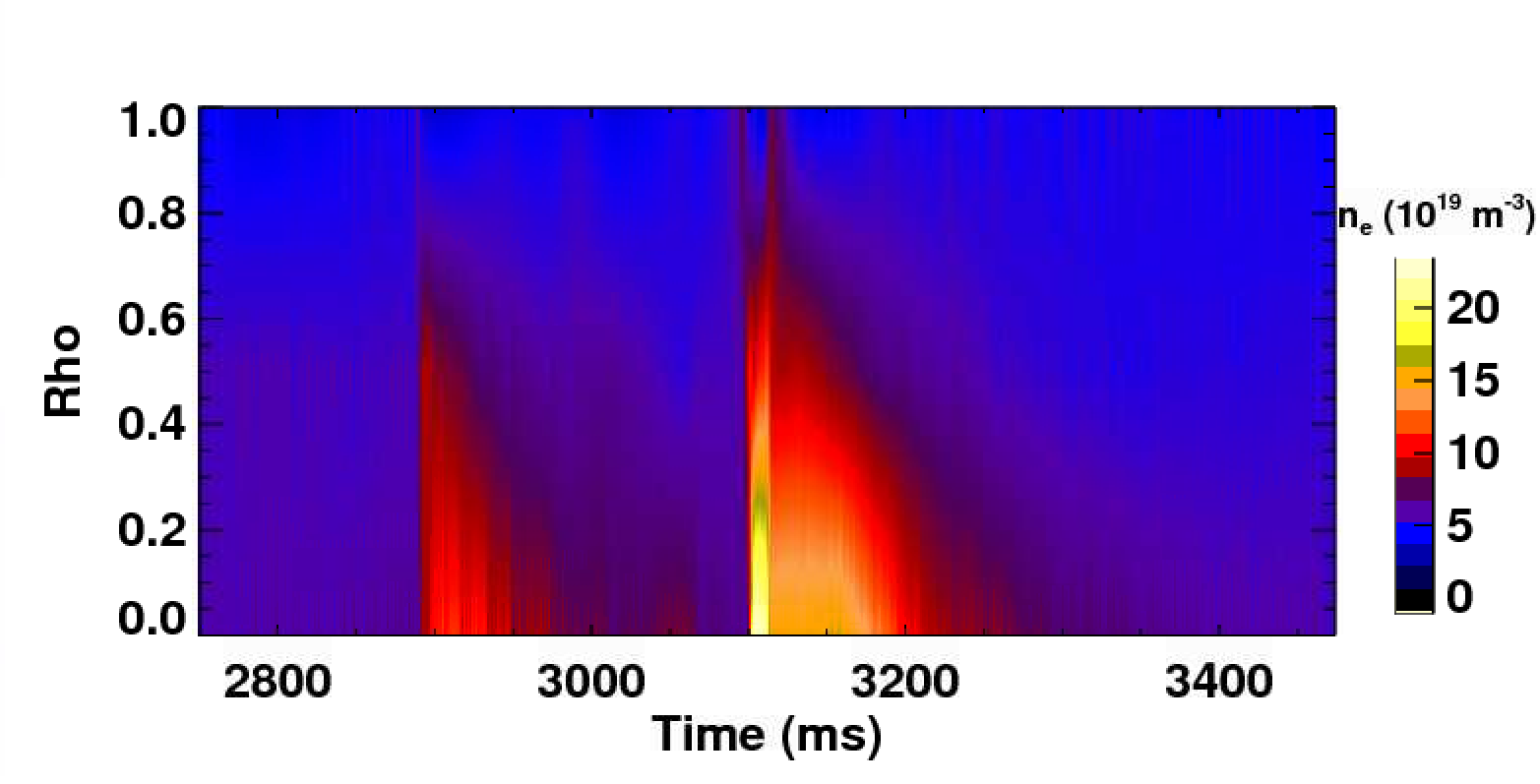}
        \caption{Reconstructed $n_e(\rho,t)$ from four CO$_2$ interferometer chords showing localized density increases from two pellet ablation events.}
        \label{fig:ne_rhot}
    \end{subfigure}
    \hfill
    \begin{subfigure}[b]{0.47\textwidth}
        \centering
        \includegraphics[width=\textwidth]{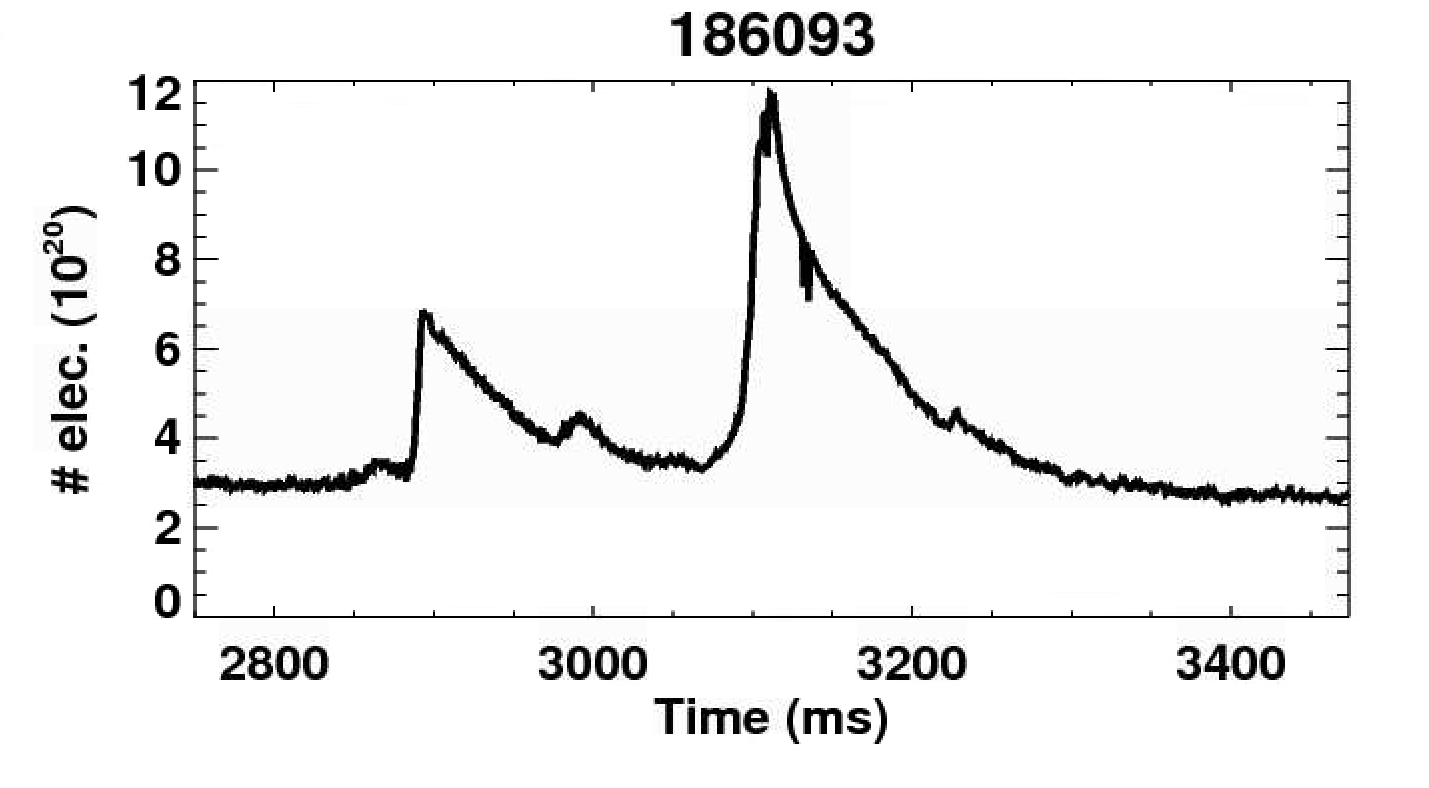}
        \caption{Excess line-integrated electron content as a function of time obtained from inversion of interferometer data.}
        \label{fig:ne_time}
    \end{subfigure}

    \vspace{0.3cm}

    \begin{subfigure}[b]{0.40\textwidth}
        \centering
        \includegraphics[width=\textwidth]{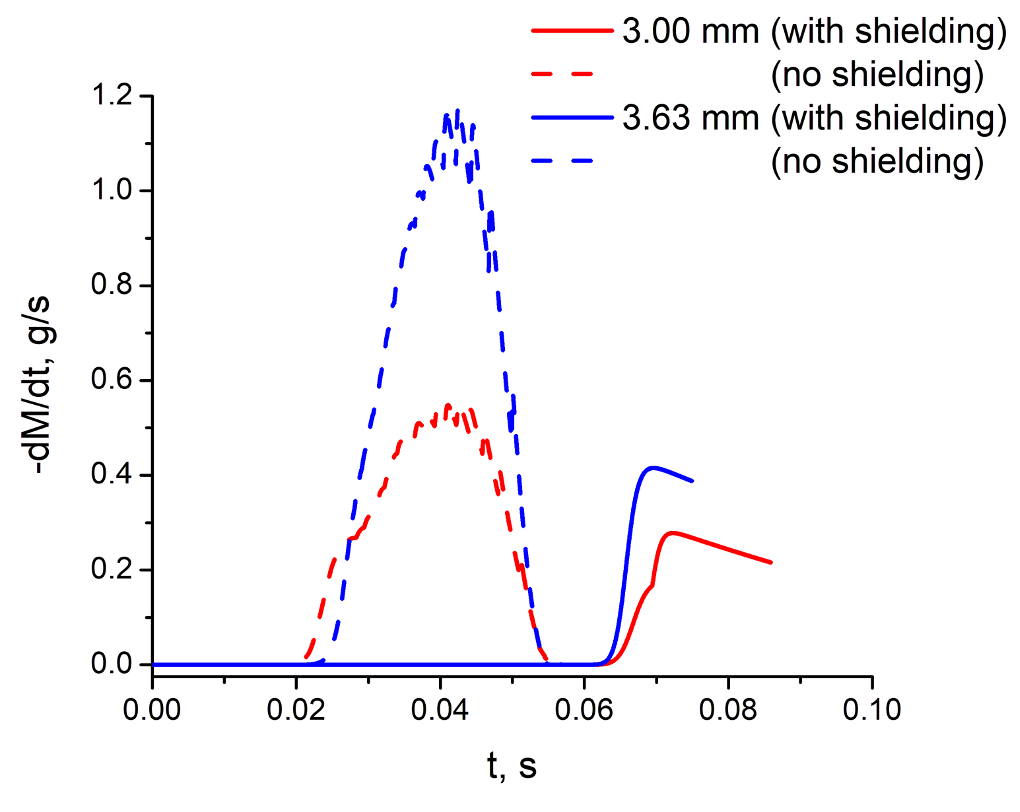}
        \caption{DUSTT-predicted pellet mass loss rates for different pellet sizes, with and without radiation-cloud shielding.}
        \label{fig:dustt_ablation}
    \end{subfigure}

    \caption{Interferometer-based evaluation of pellet ablation and comparison with DUSTT modeling.}
    \label{fig:pellet_ne_ablation}
\end{figure}

The ablation of the carbon pellets is independently assessed using CO$_2$ laser interferometer measurements of line-integrated electron density \cite{Carlstrom1988RSI}. In discharge~186093, one horizontal and three vertical chords intersect the pellet trajectory. By inverting the chord signals, a time-dependent radial density profile $n_e(\rho,t)$ is reconstructed, as shown in Fig.~\ref{fig:ne_rhot}. Two localized increases in core density are visible, corresponding to the ablation of two successive pellets launched into the same discharge. Integrating $n_e(\rho,t)$ over radius yields the excess electron content as a function of time, shown in Fig.~\ref{fig:ne_time}. Although single-pellet injections were commonly used in this study, discharge~186093 contained two well-separated pellet injections and is shown here as a representative example of the interferometer response.

The interferometer signal provides an indirect measure of pellet ablation through the ionization of ablated carbon. However, the observed temporal evolution reflects not only the ablation process itself but also subsequent impurity transport and confinement within the plasma. Consequently, the macroscopically observed duration of the density perturbation ($\sim 200$~ms) significantly exceeds the true pellet lifetime and cannot be interpreted as a direct measurement of the instantaneous mass-loss rate. Specifically, the two successive events in discharge 186093 correspond to an initial $3.00$~mm ($\approx 11.4$~mg) pellet followed by a larger $3.63$~mm ($20.3$~mg) pellet, which directly accounts for why the second experimental peak in Fig.~\ref{fig:ne_time} reaches a substantially higher line-integrated amplitude ($\approx 12 \times 10^{20}$ electrons vs. $\approx 7 \times 10^{20}$ electrons). While an initial global estimate can be obtained by averaging the mass loss over the sharp $25\text{--}30$~ms rise time of the diagnostic signals (yielding $\approx 0.4\text{--}0.5\text{~g~s}^{-1}$ for the first pellet and $\approx 0.7\text{--}0.8\text{~g~s}^{-1}$ for the second), a more rigorous, spatially resolved mass-loss rate is extracted by correlating the synchronized camera tracking coordinates from Fig.~\ref{fig:pellet_3d} directly with the localized density perturbations in Fig.~\ref{fig:ne_rhot}. Evaluating the incremental change in the local electron inventory at specific radial locations over a narrow time increment corresponding to a single fast-camera frame ($\Delta t$) maps a dynamic, position-dependent mass-loss rate throughout the pellet's inward trajectory. This localized empirical approach effectively filters out long-term core transport effects, enabling a direct, position-by-position comparison with the spatially profiled ablation rates predicted by the simulations.

Figure~\ref{fig:dustt_ablation} presents UEDGE--DUSTT predictions of pellet mass-loss rates for representative pellet sizes, with and without radiation-cloud shielding. Simulations without shielding predict rapid ablation and short pellet lifetimes, whereas inclusion of radiation-cloud shielding substantially reduces the peak mass-loss rate and extends pellet survival. Specifically, the modeling demonstrates that a $3.00\text{~mm}$ pellet reaches a maximum mass-loss rate of $0.4\text{~g~s}^{-1}$ without shielding compared to $0.22\text{~g~s}^{-1}$ with shielding, while simultaneously delaying the onset of active ablation from $0.02\text{~s}$ to $0.06\text{~s}$ to significantly prolong particle survival. Similar scaling trends are observed for the larger pellet configuration. These numerical shielding timescales are highly consistent with the experimental observations; because the second launched pellet was physically larger, it yielded a proportionally broader ablation duration and a significantly higher peak of ionized density across the diagnostic chords before completing its trajectory. These results indicate that radiation-cloud shielding plays an important role in determining pellet ablation dynamics under the high-heat-flux conditions investigated here.

\subsection{Observation of surface spallation in glassy carbon pellets}
\label{subsec:spallation}

Spallation refers to the removal of fragments of material from a solid surface due to mechanical stress, rapid heating, or hydrodynamic forces. In the context of ablative thermal protection systems (TPS), spallation represents a failure mode because the ejected fragments do not contribute to heat absorption and result in inefficient use of material mass. The competition between ablative efficiency and structural integrity therefore dictates an optimal performance region for TPS materials. When operated outside this regime, excessive surface erosion via spallation can lead to higher mass loss for a given heat flux.

During the Galileo atmospheric entry mission, spallation processes were predicted to contribute up to $\sim$10\% of the total mass loss at the probe forebody, occurring above a spallation heat flux threshold of approximately 146~MW\,m$^{-2}$~\cite{Lundell1982}. This threshold corresponds to conditions where ablation pressure and subsurface gas release overcome the mechanical binding strength of the material, ejecting fragments from the surface rather than gradually vaporizing or oxidizing them.

The parallel heat fluxes applied to the carbon samples in the experiments presented here (Sec.~\ref{sec:rods} and \ref{subsec:pellet_imaging_modeling}) reached $30$--$40~\mathrm{MW\,m^{-2}}$ for the stationary rods near the DiMES location and up to $\sim100~\mathrm{MW\,m^{-2}}$ along pellet trajectories in the confined plasma. These values are below the nominal spallation threshold of $146~\mathrm{MW\,m^{-2}}$ discussed above; yet, spallation of glassy carbon pellets was observed in our experiments. This observed fragment ejection may involve additional effects beyond the steady-state heat-flux criterion, such as localized transient heating, thermal stress gradients, material heterogeneity, or mechanical stresses associated with rapid pellet acceleration and plasma--flow interaction.

Nevertheless, spallation-like events were observed during several pellet injections. In particular, when glassy carbon spheres were injected in discharge 186153, high-speed visible imaging revealed the expulsion of micron-scale bright fragments trailing behind the pellet, consistent with spallation ejecta rather than purely evaporative ablation. 
These observations indicate that material fragmentation can occur under conditions where the average incident heat flux remains below the nominal threshold, highlighting the need for further investigation of transient thermal stress gradients and material microstructural geometry—specifically the impermeable, brittle cross-linked network of glassy carbon versus the compliant, open-pore matrix of porous graphite—as primary drivers of structural failure.


The scikit-image Python library was used to process raw Fastcam visible light imaging data to segment significant spallation ejecta and the glassy carbon main pellet according to the following workflow \cite{Van2014}. A blurred version of the raw image was obtained via convolution with a Gaussian kernel, and this blurred image was then subtracted from the raw image to filter out background noise. The contrast of this filtered image was then enhanced by rescaling its intensity values using low and high percentiles of the filtered image as intensity bounds. Next, ejecta and pellet pixels were amplified relative to the background of this enhanced image through bandpass filtering via convolution of the intensity--enhanced image with a difference of Gaussian kernels---a standard method referred to in image processing as Difference of Gaussian (DoG) blob detection. Finally, masks of ejecta and the main pellet were obtained by thresholding the DoG image. Representative results for a Fastcam frame are shown in Figure~\ref{fig:pellet_ne_ablation}.

The primary objective of this analysis was to establish the ability to directly identify and track spallation events in high-speed camera data. The image-processing workflow successfully separates the primary pellet from detached fragments and provides a foundation for future quantitative measurements of ejecta size, velocity, and production rate. In the present work, the analysis is used qualitatively to demonstrate the occurrence of spallation during pellet ablation in tokamak plasma conditions. The ability to access this regime experimentally enables testing of a wide class of TPS materials and the simultaneous validation of thermochemical ablation and fracture-driven erosion models.



\begin{figure}[htbp]
    \centering
    \begin{subfigure}[b]{0.47\textwidth}
        \centering
        \includegraphics[width=\textwidth]{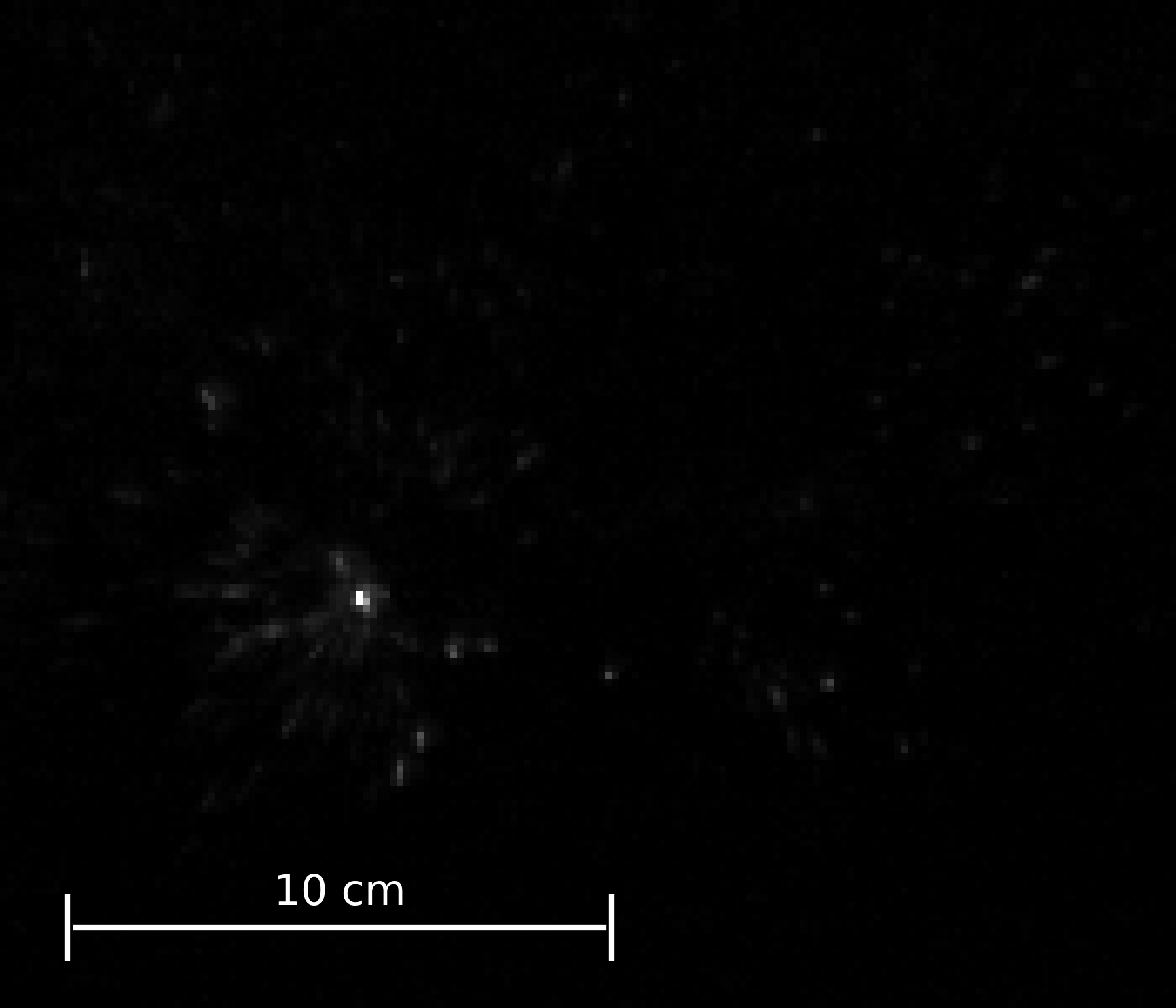}
        \caption{Raw image.}
        \label{fig:fastcam_ejecta_raw}
    \end{subfigure}
    \hfill
    \begin{subfigure}[b]{0.47\textwidth}
        \centering
        \includegraphics[width=\textwidth]{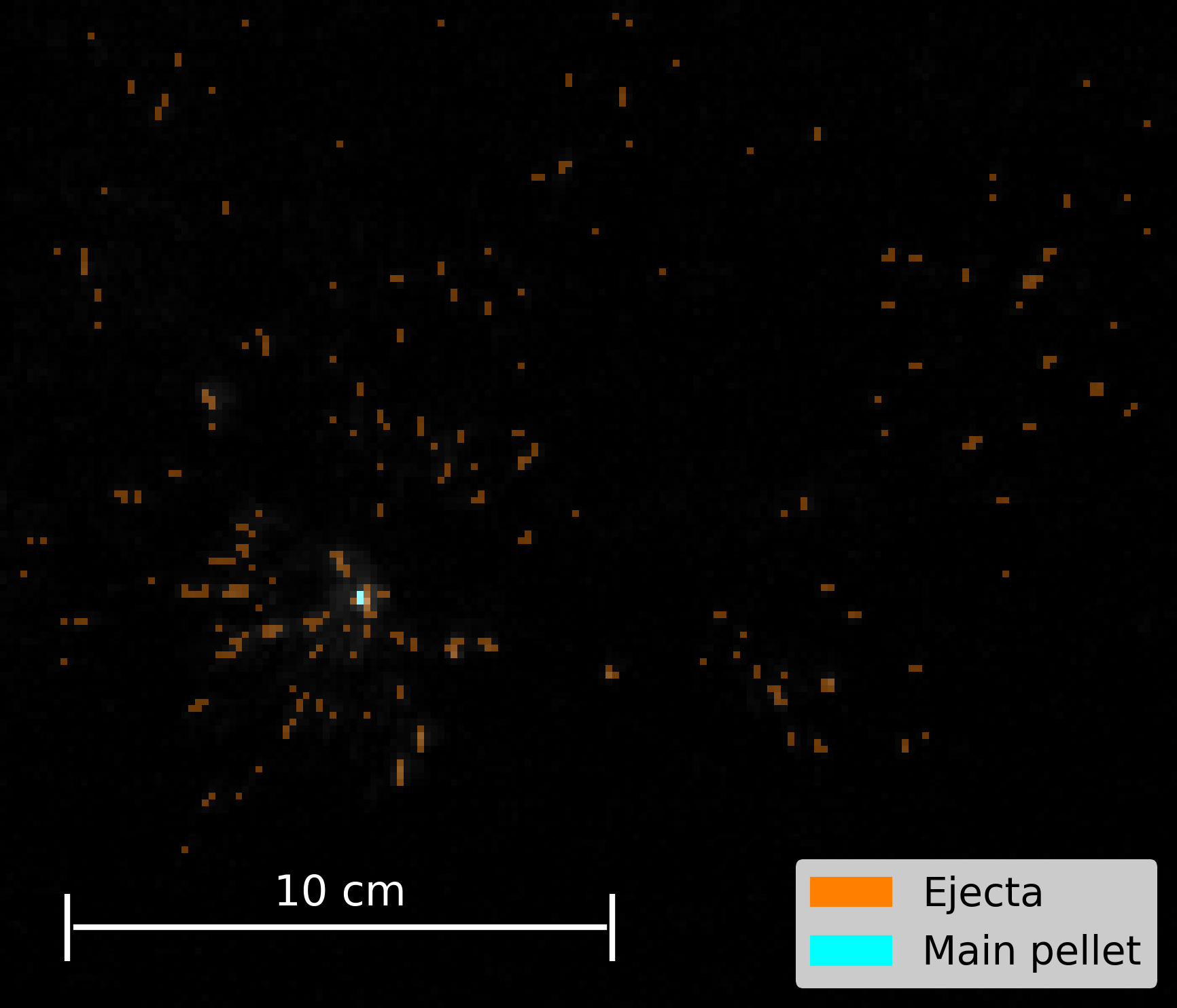}
        \caption{Raw image with ejecta and main pellet masks overlaid.}
        \label{fig:fastcam_ejecta_processed}
    \end{subfigure}

    \vspace{0.3cm}

    \caption{Visible imaging at time $t=2.6317$s of material fragment ejecta trailing a glassy-carbon pellet injected into discharge 186153. These ejecta exhibit similar morphology to surface-spallation experiments performed under extreme heating conditions in atmospheric entry studies. The scale bar represents distance along the ground and is not representative of distance between ejecta.}
    \label{fig:pellet_ne_ablation}
\end{figure}

\section{Conclusions}
\label{sec:conclusions}

This work demonstrates that the scrape-off layer (SOL) and divertor plasma in the DIII-D tokamak can serve as a unique laboratory platform for studying material response, ablation, and thermal protection system (TPS) performance under heat-flux conditions directly relevant to planetary atmospheric entries. By leveraging controlled plasma conditions, well-defined material exposures, and comprehensive multi-diagnostic measurements, we have produced the first systematic investigation of carbon ablation in a fusion environment specifically aimed at informing thermal protection system (TPS) modeling for giant-planet entries.

Two complementary ablation techniques were developed and deployed: stationary carbon rods exposed to fixed divertor strike-point heat fluxes, and millimeter-scale carbon pellets injected deep into the confined plasma. These approaches enabled characterization of both solid-surface recession under intense plasma heat flux and volumetric ablation in the plasma core, akin to the pyrolysis and char-layer removal experienced by ablators during atmospheric entry.

High-resolution 3-D profilometry, precision mass loss measurements, and spectroscopic observations show that stationary carbon rods experienced ablation consistent with parallel heat fluxes in the range of 30--40~MW\,m$^{-2}$, comparable to those predicted for the Galileo probe nose region \cite{LaubVenkatapathy2003}. The results were systematically compared against the Park and Matsuyama semi-empirical engineering models. The cylindrical and concave rod experiments were found to be in closer agreement with the Park model predictions, while the wedge-shaped rod exhibited substantially higher ablation rates and stronger geometric sensitivity. These observations suggest that Park's model captures the average ablation behavior more accurately under the present conditions, whereas additional geometry-dependent effects must be considered to explain the enhanced recession observed for the wedge-shaped samples.

Pellet experiments enabled full 3-D reconstruction of particle trajectories using synchronized visible-light and infrared imaging. The observed dynamics revealed the importance of electromagnetic ($\mathbf{j}\times\mathbf{B}$) forces for millimeter-scale particles in magnetized non-uniform plasmas---a mechanism typically negligible for micron-scale dust. Incorporation of this force into UEDGE--DUSTT simulations was essential to reproduce observed trajectories and to avoid premature wall losses seen in simplified models. Furthermore, line-integrated density measurements from the CO$_2$ interferometer allowed inference of ablated carbon mass on millisecond timescales, enabling initial validation of carbon pellet ablation predictions under fusion-edge plasma conditions.

A key finding of this work is that \textit{spallation} - fragmentation and ejection of carbon fragments under high mechanical and thermal stresses - was observed when using brittle glassy carbon pellets. 
This behavior occurred despite local parallel heat fluxes remaining below the nominal 146~MW\,m$^{-2}$ steady-state spallation threshold previously predicted for Galileo’s heat shield \cite{Lundell1982}, demonstrating that transient thermal stress gradients and tight material microstructures can lower the onset threshold for mechanical fracturing. This result underscores the unique value of tokamak-based ablation studies for capturing real TPS failure modes and operational limits that cannot be fully replicated in traditional ground-based testing.

Overall, these results establish a new pathway for improving predictive TPS ablation models by integrating:
\begin{itemize}
\item controlled plasma experiments under flight-relevant heat loads and geometries,
\item real-time imaging and spectroscopy of ablation processes,
\item validated transport and ablation simulations, and
\item material-specific erosion response data.
\end{itemize}

This study provides the first significant feedback from fusion-plasma experiments into TPS engineering models originally developed for the Galileo mission and future giant-planet probes. Continued refinement of these experiments - including improved automated image-based tracking, in-situ recession diagnostics, and advanced radiation-shielding physics - will enable increasingly accurate predictions of material ablation and spallation for next-generation outer-planet exploration missions.
\section{Acknowledgement}
This material is based upon work supported by the U.S. Department of Energy, Office of Science, Office of Fusion Energy Sciences, using the DIII-D National Fusion Facility, a DOE Office of Science user facility, under Awards DE-FG02-05ER54809, DE-FC02-04ER54698, DE-SC0026433, DE-SC0021338, DE-SC0021620, DE-AC52-07NA27344, DE-AC02-09CH11466, DE-AC05-00OR22725, DE-SC0014264, DE-SC0023061, DE-SC0024547, and the National Science Foundation award NSF-PHY-2440328.


\begin{thebibliography}{99}

\bibitem{AllenEggers1958}
H.~J. Allen and A.~J. Eggers, Jr.
\newblock A study of the motion and aerodynamic heating of ballistic missiles entering the Earth's atmosphere at high supersonic speeds.
\newblock NACA Report 1381, National Advisory Committee for Aeronautics, 1958.
\newblock \url{https://ntrs.nasa.gov/citations/19930091020}.

\bibitem{Cutts2005EntryProbes}
J.~A. Cutts, J.~Arnold, E.~Venkatapathy, E.~Kolawa, M.~Munk, P.~Wercinski, and B.~Laub.
\newblock Technology for entry probes.
\newblock In {\em 2nd International Planetary Probe Workshop}, NASA NTRS Document ID 20070014622, Apr. 2005.
\newblock \url{https://ntrs.nasa.gov/citations/20070014622}.

\bibitem{Tauber1991Apollo}
M.~E. Tauber and K.~Sutton.
\newblock Stagnation-point radiative heating relations for Earth and Mars entries.
\newblock {\em Journal of Spacecraft and Rockets}, 28(1):40--42, 1991.
\newblock doi: \url{https://doi.org/10.2514/3.26206}.

\bibitem{Liu2010StardustRadiation}
Y.~Liu, D.~Prabhu, K.~A. Trumble, D.~Saunders, and P.~Jenniskens.
\newblock Radiation modeling for the reentry of the Stardust sample return capsule.
\newblock {\em Journal of Spacecraft and Rockets}, 47(5):741--752, Sept.--Oct. 2010.
\newblock doi: \url{https://doi.org/10.2514/1.37813}.

\bibitem{Moss1982}
J.~N. Moss and A.~L. Simmonds.
\newblock Galileo probe forebody flowfield predictions during Jupiter entry.
\newblock In {\em 3rd Joint Thermophysics, Fluids, Plasma and Heat Transfer Conference}, AIAA Paper 1982-0874, June 1982.
\newblock doi: \url{https://doi.org/10.2514/6.1982-874}.

\bibitem{CabreraWest2025PioneerVenus}
J.~V.~V. Cabrera and T.~K. West.
\newblock Pioneer Venus large probe stagnation point entry heating with coupled ablation.
\newblock {\em Journal of Spacecraft and Rockets}, 63(2):360--368, 2026.
\newblock doi: \url{https://doi.org/10.2514/1.A36431}.

\bibitem{LaubVenkatapathy2003}
B.~Laub and E.~Venkatapathy.
\newblock Thermal protection system technology and facility needs for demanding future planetary missions.
\newblock In {\em Proceedings of the International Workshop on Planetary Probe Atmospheric Entry and Descent Trajectory Analysis and Science}, ESA Special Publication SP-544, pages 239--247. European Space Agency, 2004.

\bibitem{matsuyama2005numerical}
S.~Matsuyama, N.~Ohnishi, A.~Sasoh, and K.~Sawada.
\newblock Numerical simulation of Galileo probe entry flowfield with radiation and ablation.
\newblock {\em Journal of Thermophysics and Heat Transfer}, 19(1):28--35, 2005.
\newblock doi: \url{https://doi.org/10.2514/1.10264}.

\bibitem{Milos1999}
F.~S. Milos, Y.~K. Chen, T.~H. Squire, and R.~A. Brewer.
\newblock Analysis of Galileo probe heatshield ablation and temperature data.
\newblock {\em Journal of Spacecraft and Rockets}, 36(3):298--306, 1999.
\newblock doi: \url{https://doi.org/10.2514/2.3465}.

\bibitem{park2009stagnation}
C.~Park.
\newblock Stagnation-region heating environment of the Galileo probe.
\newblock {\em Journal of Thermophysics and Heat Transfer}, 23(3):417--424, 2009.
\newblock doi: \url{https://doi.org/10.2514/1.38712}.

\bibitem{Luxon2002}
J.~L. Luxon.
\newblock A design retrospective of the DIII-D tokamak.
\newblock {\em Nuclear Fusion}, 42(5):614--633, 2002.
\newblock doi: \url{https://doi.org/10.1088/0029-5515/42/5/313}.

\bibitem{Wong1998DiMES}
C.~P.~C. Wong, D.~G. Whyte, R.~J. Bastasz, J.~Brooks, W.~P. West, and W.~R. Wampler.
\newblock Divertor materials evaluation system (DiMES).
\newblock {\em Journal of Nuclear Materials}, 258--263:433--439, 1998.
\newblock doi: \url{https://doi.org/10.1016/S0022-3115(98)00407-3}.

\bibitem{Rudakov2017}
D.~L. Rudakov, T.~Abrams, R.~Ding, H.~Y. Guo, P.~C. Stangeby, W.~R. Wampler, J.~A. Boedo, A.~Briesemeister, J.~N. Brooks, D.~A. Buchenauer et~al.
\newblock DiMES PMI research at DIII-D in support of ITER and beyond.
\newblock {\em Fusion Engineering and Design}, 124:196--201, 2017.
\newblock doi: \url{https://doi.org/10.1016/j.fusengdes.2017.03.007}.

\bibitem{Moyer2018RSI}
R.~A. Moyer, I.~Bykov, D.~M. Orlov, T.~E. Evans, J.~S. Lee, A.~M. Teklu, M.~E. Fenstermacher, M.~Makowski, C.~J. Lasnier, H.~Q. Wang et~al.
\newblock Imaging divertor strike point splitting in RMP ELM suppression experiments in the DIII-D tokamak.
\newblock {\em Review of Scientific Instruments}, 89(10):10E106, 2018.
\newblock doi: \url{https://doi.org/10.1063/1.5038350}.

\bibitem{Lasnier2014RSI}
C.~J. Lasnier, S.~L. Allen, R.~E. Ellis, M.~E. Fenstermacher, A.~G. McLean, W.~H. Meyer, K.~Morris, L.~G. Seppala, K.~Crabtree, and M.~A. Van~Zeeland.
\newblock Wide-angle ITER-prototype tangential infrared and visible viewing system for DIII-D.
\newblock {\em Review of Scientific Instruments}, 85(11):11D855, 2014.
\newblock doi: \url{https://doi.org/10.1063/1.4892897}.

\bibitem{brooks_howald_klepper_west_1992}
N.~H. Brooks, A.~Howald, K.~Klepper, and P.~West.
\newblock Multichord spectroscopy of the DIII-D divertor region.
\newblock {\em Review of Scientific Instruments}, 63(10):5167--5169, 1992.
\newblock doi: \url{https://doi.org/10.1063/1.1143469}.

\bibitem{Carlstrom1995DTS}
T.~N. Carlstrom, J.~H. Foote, D.~G. Nilson, and B.~W. Rice.
\newblock Design of the divertor Thomson scattering system on DIII-D.
\newblock {\em Review of Scientific Instruments}, 66(1):493--495, 1995.
\newblock doi: \url{https://doi.org/10.1063/1.1146534}.

\bibitem{Watkins1992RecipProbe}
J.~G. Watkins, R.~A. Moyer, J.~W. Cuthbertson, D.~A. Buchenauer, T.~N. Carlstrom, D.~N. Hill, and M.~Ulrickson.
\newblock Reciprocating and fixed probe measurements of density and temperature in the DIII-D divertor.
\newblock {\em Journal of Nuclear Materials}, 241--243:645--649, 1997.
\newblock doi: \url{https://doi.org/10.1016/S0022-3115(97)80115-8}.

\bibitem{Buchenauer1990FixedProbes}
D.~Buchenauer, W.~L. Hsu, J.~P. Smith, and D.~N. Hill.
\newblock Langmuir probe array for the DIII-D divertor.
\newblock {\em Review of Scientific Instruments}, 61(10):2873--2875, 1990.
\newblock doi: \url{https://doi.org/10.1063/1.1141811}.

\bibitem{Tauber1999NASA}
M.~E. Tauber, P.~Wercinski, L.~Yang, and Y.-K. Chen.
\newblock A fast code for Jupiter atmospheric entry analysis.
\newblock NASA Technical Memorandum NASA/TM-1999-208796, NASA Ames Research Center, Moffett Field, CA, Sept. 1999.

\bibitem{Erb2020GalileoHeating}
A.~J. Erb, T.~K. West, and C.~O. Johnston.
\newblock Investigation of Galileo probe entry heating with coupled radiation and ablation.
\newblock NASA Preprint NF1676L-35423, NASA Langley Research Center, Jan. 2020.
\newblock NASA Document ID 20200003199; preprint submitted to {\em AIAA Journal of Spacecraft and Rockets}.

\bibitem{Watkins2001JNM}
J.~G. Watkins, P.~C. Stangeby, J.~A. Boedo, T.~N. Carlstrom, C.~J. Lasnier, R.~A. Moyer, D.~L. Rudakov, and D.~G. Whyte.
\newblock Comparison of Langmuir probe and Thomson scattering measurements in DIII-D.
\newblock {\em Journal of Nuclear Materials}, 290--293:778--782, 2001.
\newblock doi: \url{https://doi.org/10.1016/S0022-3115(00)00534-1}.

\bibitem{donovan2013experimental}
D.~Donovan, D.~Buchenauer, J.~Watkins, A.~Leonard, C.~Wong, M.~Schaffer, D.~Rudakov, C.~Lasnier, and P.~Stangeby.
\newblock Experimental measurements of the particle flux and sheath power transmission factor profiles in the divertor of DIII-D.
\newblock {\em Journal of Nuclear Materials}, 438:S467--S471, 2013.
\newblock doi: \url{https://doi.org/10.1016/j.jnucmat.2013.01.095}.

\bibitem{Smirnov2007}
R.~D. Smirnov, A.~Yu. Pigarov, M.~Rosenberg, S.~I. Krasheninnikov, and D.~A. Mendis.
\newblock Modelling of dynamics and transport of carbon dust particles in tokamaks.
\newblock {\em Plasma Physics and Controlled Fusion}, 49(4):347--371, 2007.
\newblock doi: \url{https://doi.org/10.1088/0741-3335/49/4/001}.

\bibitem{Orlov2021IMECE}
D.~M. Orlov, M.~O. Hanson, J.~Escalera, H.~Taheri, C.~N. Villareal, D.~M. Zubovic, I.~Bykov, E.~G. Kostadinova, D.~L. Rudakov, and M.~Ghazinejad.
\newblock Design and testing of DiMES carbon ablation rods in the DIII-D tokamak.
\newblock In {\em ASME International Mechanical Engineering Congress and Exposition}, Vol. 4: Advances in Aerospace Technology, V004T04A038. ASME, 2021.
\newblock doi: \url{https://doi.org/10.1115/IMECE2021-73326}.

\bibitem{Eich2013}
T.~Eich, A.~W. Leonard, R.~A. Pitts, W.~Fundamenski, R.~J. Goldston, T.~K. Gray, A.~Herrmann, A.~Kirk, A.~Kallenbach, O.~Kardaun et~al.
\newblock Scaling of the tokamak near the scrape-off layer H-mode power width and implications for ITER.
\newblock {\em Nuclear Fusion}, 53(9):093031, 2013.
\newblock doi: \url{https://doi.org/10.1088/0029-5515/53/9/093031}.

\bibitem{Herzberg1950}
G.~Herzberg.
\newblock {\em Molecular Spectra and Molecular Structure. Vol. I: Spectra of Diatomic Molecules}.
\newblock D. Van Nostrand, 1950.

\bibitem{Wernitz2011AIAA}
R.~Wernitz, C.~Eichhorn, G.~Herdrich, S.~L{\"o}hle, S.~Fasoulas, and H.~P. R{\"o}ser.
\newblock Plasma wind tunnel investigation of European ablators in air using emission spectroscopy.
\newblock In {\em 42nd AIAA Thermophysics Conference}, Honolulu, Hawaii, AIAA Paper 2011-3761, June 2011.
\newblock doi: \url{https://doi.org/10.2514/6.2011-3761}.

\bibitem{McLean2009PhD}
A.~G. McLean.
\newblock {\em Quantification of Chemical Erosion in the Divertor of the DIII-D Tokamak}.
\newblock Ph.D. thesis, University of Toronto, Toronto, Ontario, Canada, Nov. 2009.

\bibitem{SiegelHowell}
R.~Siegel and J.~R. Howell.
\newblock {\em Thermal Radiation Heat Transfer}.
\newblock 4th edition, Taylor \& Francis, 2002.

\bibitem{Rognlien1992}
T.~D. Rognlien, J.~L. Milovich, M.~E. Rensink, and G.~D. Porter.
\newblock A fully implicit, time dependent 2-D fluid code for modeling tokamak edge plasmas.
\newblock {\em Journal of Nuclear Materials}, 196--198:347--351, 1992.
\newblock doi: \url{https://doi.org/10.1016/S0022-3115(06)80058-9}.

\bibitem{Pigarov2005}
A.~Yu. Pigarov, S.~I. Krasheninnikov, T.~K. Soboleva, and T.~D. Rognlien.
\newblock Dust-particle transport in tokamak edge plasmas.
\newblock {\em Physics of Plasmas}, 12(12):122508, 2005.
\newblock doi: \url{https://doi.org/10.1063/1.2145157}.

\bibitem{Krasheninnikov2009}
S.~I. Krasheninnikov and R.~D. Smirnov.
\newblock On interaction of large dust grains with fusion plasma.
\newblock {\em Physics of Plasmas}, 16(11):114501, 2009.
\newblock doi: \url{https://doi.org/10.1063/1.3262505}.

\bibitem{Brown2014}
B.~T. Brown, R.~D. Smirnov, and S.~I. Krasheninnikov.
\newblock On vapor shielding of dust grains of iron, molybdenum, and tungsten in fusion plasmas.
\newblock {\em Physics of Plasmas}, 21(2):024501, 2014.
\newblock doi: \url{https://doi.org/10.1063/1.4866599}.

\bibitem{Krasheninnikov2014}
S.~I. Krasheninnikov, R.~D. Smirnov, and D.~L. Rudakov.
\newblock Dust in magnetic fusion devices.
\newblock {\em Plasma Physics and Controlled Fusion}, 53(8):083001, 2011.
\newblock doi: \url{https://doi.org/10.1088/0741-3335/53/8/083001}.

\bibitem{Marenkov2014}
E.~D. Marenkov and S.~I. Krasheninnikov.
\newblock Ablation of high-Z material dust grains in edge plasmas of magnetic fusion devices.
\newblock {\em Physics of Plasmas}, 21(12):123701, 2014.
\newblock doi: \url{https://doi.org/10.1063/1.4903333}.

\bibitem{Smirnov2020}
R.~D. Smirnov and S.~I. Krasheninnikov.
\newblock Time-dependent modeling of dust outburst into tokamak divertor plasma.
\newblock {\em Physics of Plasmas}, 27(8):082509, 2020.
\newblock doi: \url{https://doi.org/10.1063/5.0009767}.

\bibitem{Carlstrom1988RSI}
T.~N. Carlstrom, D.~R. Ahlgren, and J.~Crosbie.
\newblock Real-time, vibration-compensated CO$_2$ interferometer operation on the DIII-D tokamak.
\newblock {\em Review of Scientific Instruments}, 59(7):1063--1066, 1988.
\newblock doi: \url{https://doi.org/10.1063/1.1139726}.

\bibitem{Lundell1982}
J.~H. Lundell.
\newblock Spallation of the Galileo probe heat shield.
\newblock NASA NTRS Report 19820052717, NASA, 1982; also published as AIAA Paper 82-0852.
\newblock doi: \url{https://doi.org/10.2514/6.1982-852}.

\bibitem{Van2014}
S.~van~der~Walt, J.~L. Sch{\"o}nberger, J.~Nunez-Iglesias, F.~Boulogne, J.~D. Warner, N.~Yager, E.~Gouillart, T.~Yu, and the scikit-image contributors.
\newblock scikit-image: Image processing in Python.
\newblock {\em PeerJ}, 2:e453, 2014.
\newblock doi: \url{https://doi.org/10.7717/peerj.453}.

\end{thebibliography}

\newpage
\section*{Disclaimer}
This report was prepared as an account of work sponsored by an agency of the United States Government. Neither the United States Government nor any agency thereof, nor any of their employees, makes any warranty, express or implied, or assumes any legal liability or responsibility for the accuracy, completeness, or usefulness of any information, apparatus, product, or process disclosed, or represents that its use would not infringe privately owned rights. Reference herein to any specific commercial product, process, or service by trade name, trademark, manufacturer, or otherwise does not necessarily constitute or imply its endorsement, recommendation, or favoring by the United States Government or any agency thereof. The views and opinions of authors expressed herein do not necessarily state or reflect those of the United States Government or any agency thereof.
\end{document}